\documentclass[prb,twocolumn,showpacs,preprintnumbers,amsmath,amssymb]{revtex4}
\usepackage{graphicx}% Include figure files
\usepackage{dcolumn}% Align table columns on decimal point
\usepackage{bm}% bold math
\usepackage{color}
\usepackage{soul}
\usepackage{multirow}
\newcommand{\mycite}[1]{\scalebox{1.4}[1.4]{\raisebox{-0.80ex}{\cite{#1}}}}
\usepackage[section]{placeins}
\begin{document}

 \preprint{APS/123-QED}
\title{ Extended states in one-dimensional aperiodic lattices with linearly varying patches}

%\title{Modeling and control epidemic spreading of an SIR model in a residential university environment}

\author{Longyan Gong  $^{1,2}$}
\thanks{Email address: lygong@njupt.edu.cn.}

\affiliation{
 $^{1}$College of Science, Nanjing University of Posts and Telecommunications, Nanjing, 210003, China\\
 $^{2}$Jiangsu Provincial Engineering Research Center of Low Dimensional Physics and New Energy, Nanjing, 210003, China\\
 }
\date{today}
\begin{abstract}
We introduce a family of 1D aperiodic tight-binding models with linearly varying patches of A-type
sites with on-site energies $\epsilon_A=0$ connected by single B-type sites with $\epsilon_B=W$. We analytically show such structures have strong spatial correlations. We theoretically find states are extended at resonance levels in the vicinity of $E^\kappa_M=-2\cos\frac{\kappa\pi}{M}$ if they are allowed energies, where $M=md$ are the size differences of patches, $d$ is the variation rate of patch sizes, $m\in \mathcal{N_{+}}$ and $\kappa=1,2,\cdots,M-1$. Related delocalization-localization transitions are explored. Numerical evidences are in excellent quantitative agreement with theoretical predictions.
\end{abstract}

%71.23.An Theories and models; localized states
%72.15.Rn Localization effects (Anderson or weak localization)
%71.30.+h Metal-insulator transitions and other electronic transitions
%71.23.Ft Quasicrystals
%-------------------------------
\maketitle
\section{Introduction}\label{Sec1}
For a long time, the localization phenomenon attracts a lot of attention in condensed matter physics~\cite{LA09,BH21}. In uncorrelated disorder potentials, Anderson transitions of noninteracting electrons can happen in 3D systems as potentials increase~\cite{AN58,AB10,EV08}. Such transitions are sometimes referred to as metal-insulator transitions or delocalization-localization transitions. Periodic potentials have translational symmetry, and according to Bloch's theory, all states are extended. In contrast, aperiodic potentials have no translational symmetry but have strong spatial correlations, which is clearly distinguished from the disordered and periodic ones~\cite{MA09}. For instance, states in the  Aubry-Andr\'{e}-Harper model may be extended, critical, or localized, which depend on potential strength~\cite{HA55,AU80}. There are extended critical states in the Fibonacci lattices~\cite{MA96} and there may be extended states in the Thue-Morse ones~\cite{RY92,CH95}. Thus states in aperiodic structures exhibit intermediate properties. In practice, aperiodic structures can provide an inspiring guide to design devices in optoelectronics, optical communication applications and others~\cite{MA21}. So it is interesting to propose novel aperiodic structures with remarkable properties.

At the same time, it is a challenging and important problem to understand
the nature of quantum states~\cite{BR92}, i.e., whether states are extended or localized. Disorder can induce localized states and inhibit electronic, vibrational, and transport properties~\cite{AB10}. However, not all states in 1D disordered systems are localized. For example, multiple-resonance necklace states are typical quasi-extended states, which are formed due to the coupling of many nearly degenerate localized states that are centered at different parts of a chain~\cite{PE87,PE94}. These localized modes are strung together like beards around the necklace, so such states are called necklace states. They can improve the electronic transport properties. A different approach to creating localized states is in periodic systems with a flat-band spectrum~\cite{LE18}. Due to internal symmetries or fine-tuned coupling, flat-band states are perfectly localized to several lattice sites, leading to compact localized states. In such systems, the disorder can induce delocalization, i.e., a transition from an insulating to a metallic phase, dubbed inverse Anderson transition~\cite{GO06}. More interestingly, resonance non-scattered states have also been found in a 1D random-dimer tight-binding model~\cite{DU90}, where A-type and B-type sites are randomly distributed and one component appears in pairs. There are short-range correlations in its on-site potentials. Since then, resonance states are found in similar models, for instance, random trimer~\cite{GI93}, random dimer-trimer~\cite{FA97} and random n-mer ones~\cite{EV93,GO16,IZ95}.

Recently, Bykov \emph{et al.} proposed guided-mode resonant gratings with linearly varying periods~\cite{BY22}. These structures can exhibit resonance reflectance peaks with the spatial position depending on the incident wavelength, so such gratings can be used as novel optical filters. It is worth extending it to other fields. Very recently, Citrin proposed quadratic superlattices and found extended states~\cite{CI23}. In fact, it is a specific linearly-varying-period structure. However, the mechanism of the presence of extended states should be further explained.

Inspired by the above-mentioned, we propose a family of 1D aperiodic lattices with linearly varying periods. We analytically show such structures have strong spatial correlations. With a tight-binding model, we theoretically find extended states at resonance energies and their underlying mechanism. The state localization properties are also intensively certified by numerical evidences.

\section{Model}\label{Sec2}

\begin{figure*}[!htbp]%fig1
\centering
 \includegraphics[width=6.0in]{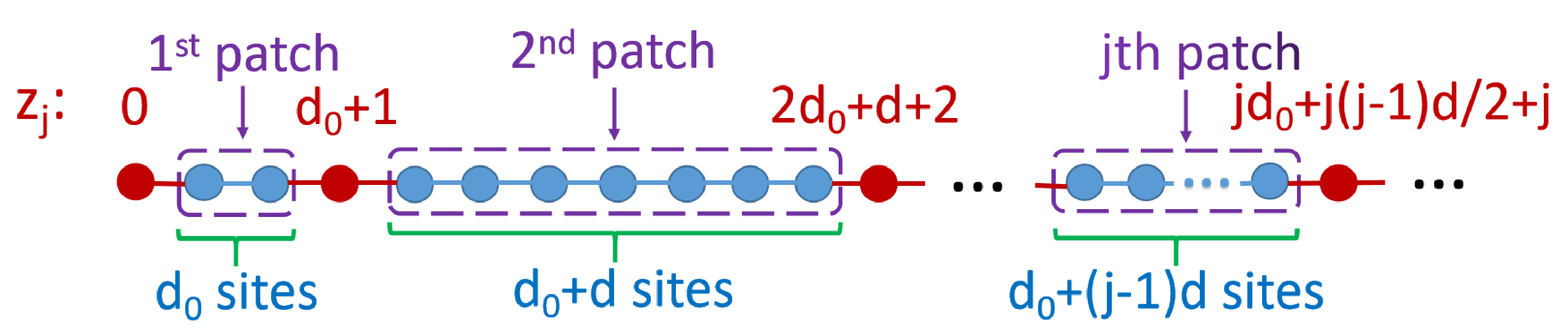}
 \caption{Schematic diagram of 1D aperiodic lattices. The $j$th linearly varying patch has $s_j=d_0+(j-1)d$ A-type sites (blue solid circles), where $d_0$ is the size of first patch and $d$ is the variation rate of patch sizes; the inlaid single B-type sites (red solid circles) link these patches with positions $Z_j=jd_0+j(j-1)d/2+j$. Here, $d_0=2$ and $d=5$ are as examples.}\label{Fig1}
\end{figure*}

The family of 1D aperiodic structure is sketched in Fig.~\ref{Fig1}, where the distance between two nearest sites is set to the unit. The $j$th patch has $s_j=d_0+(j-1)d$ A-type sites, and we call it $s_j$-mer. Here, $d_0$ is the size of first patch, $d$ is the variation rate of patch sizes, and $j=1,2,\cdots,j_m$. They are linearly varying patches (LVPs)~\cite{BY22}. The  inlaid single B-type sites link these LVPs with positions $Z_j=j(j+1)d/2+j$. The basic length $d$ defines a family of structures. The structure is periodical if $d=0$. It becomes the model proposed in Ref.~\mycite{CI23} if $d_0=0$ and $d=2$. Further, these $s_j$-mer are arranged in order of increasing size and there are long-range correlations in structures (seen the next section), which is different from that in the random-dimer model~\cite{DU90} as well as its variants~\cite{GI93,FA97,EV93,GO16,IZ95} having short-range correlations.

For an electron moving in such structures, the nearest-neighbour tight-binding Hamiltonian can be written as
\begin{equation}
H=\sum\limits_{n=0}^{N-1}\varepsilon_n |n
\rangle\langle n|-
  t\sum\limits_{n=0}^{N-2} (|n \rangle\langle n+1|+|n+1 \rangle\langle n|),\label{EQ1}
\end{equation}
where $|n\rangle=c^{\dag}|0\rangle$, $c^{\dag}$ is the creation operator, $\varepsilon_n$ is the on-site potential, and $t$ is the nearest-neighbor hopping integral. The system size $N=Z_{j_m}+1$. In Fig.~\ref{Fig1}, for simplicity, we suppose the A-type sites have zero potentials and the B-type sites have constant potentials with strength $W$, i.e.,
\begin{eqnarray}
\varepsilon_n=\left\{
\begin{array}
{r@{}l}$$W,$~~~$n=Z_j~\textrm{and}~j=0,1,2,\cdots,j_m,\\
       $$0,$~~~$\textrm{otherwise}.~~~~~~~~~~~~~~~~~~~~~~~~~~~
\end{array}\right.\label{EQ2}
\end{eqnarray}
At $W=0$, eigenstates can be expressed by $|\Psi_\beta\rangle=\sum\limits_{n=0}^{N-1}\psi_\beta(n)|n \rangle=\sum\limits_{n=0}^{N-1} \sqrt{\frac{2}{N+1}}\sin[\frac{\beta{(n+1)}\pi}{N+1}]|n\rangle$ with $\beta=1,2,\dots,N$, where all states are extended~\cite{KU99,TO19}.

\section{Results}\label{Sec3}
We will study structure factors of the 1D aperiodic lattices shown in Fig.~\ref{Fig1}, energy spectrum properties of the Hamiltonian in Eq.(\ref{EQ1}), state localization properties, and the effect of randomness. In the following, results for $d=5$ along with $d_0=2$ are presented.

\begin{figure}[!htbp]%fig2
 \includegraphics[width=1.6in]{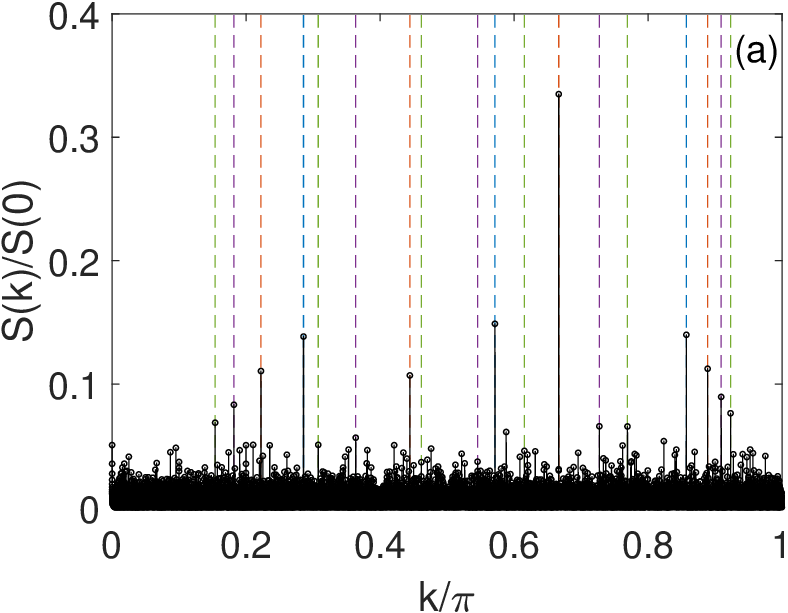}
 ~~~\includegraphics[width=1.6in]{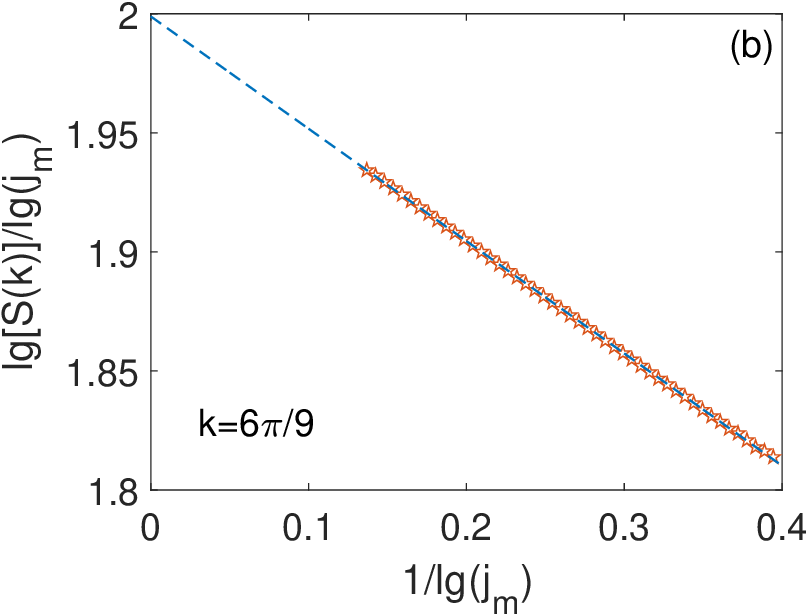}
 \caption{(a) Structure factors $S(k)$ versus wave vectors $k$ at system size $N=107, 227$ ($j_m=207$). (b) Finite-size scaling of $S(k)$ at $k=6\pi/9$. In (a), the vertical dashed lines correspond to $k=\frac{2}{L}\widetilde{k}\pi$ with $\widetilde{k}=1,2,\cdots,[L/2]$,
 and $L=7$ (blue), $9$ (green), $11$ (red) and $13$ (purple), respectively. In (b), the dashed line is linearly fitted to corresponding data.}\label{Fig2}
\end{figure}
\subsection{Structure factors}\label{Sec31}
For the structure in Fig.~\ref{Fig1}, structure factors are defined by
\begin{equation}
S(k)=\big{|}\sum_{j=0}^{j_m}\exp(ikZ_j)\big{|}^2, \label{EQ3}
\end{equation}
where $k$ are wave vectors and $i=\sqrt{-1}$.
They directly relate to the results of x-ray and neutron-diffraction experiments~\cite{MA09,MA21}. They also underly behaviors observed in the electronic, vibrational, and transport properties. We set $k=\frac{2}{L}\widetilde{k}\pi$, where $L$ is an integer and $\widetilde{k}=1,2,\cdots,[L/2]$.  At $L=7, 9, 11$ and $13$,
we get
\begin{equation}
S(k)=S_Lj_m^\alpha, \label{EQ33}
\end{equation}
where $S_7=1/7$, $S_{11}=1/11$ and $S_{13}=1/13$; $S_9=1/3$ at $\widetilde{k}=3$, and $S_9=1/9$ at $\widetilde{k}=1,2$ and $4$, respectively; and the scaling exponent $\alpha=2$ (seen Appendix A). They agree with the numerical results plotted in Figs.~\ref{Fig2}(a) and (b).
Like quasiperiodic systems~\cite{CH88,OH93}, we find there are sequences of hierarchical $\delta$-function peaks in $S(k)$, i.e., for some $k$, the factor $S_L$ is finite even at system sizes $N\to\infty$. In Ref.~\mycite{CI23}, it is found $\alpha\to2$ as $k\to0$ at $d=2$ and $d_0=0$, which agrees with our results. The exponent $\alpha=2$ has been found in periodic and some quasiperiodic structures~\cite{ME88}. So it indicates that the structure in Fig.~\ref{Fig1} is aperiodic, with strong spatial correlations. They may induce extended or critical states.

\subsection{Resonance levels}\label{Sec32}

If we only consider two patches, with the theory of trace map of transfer matrices~\cite{KO83}, we analytically show
at energies
\begin{equation}
 E^\kappa_M=-2\cos\frac{\kappa\pi}{M},\label{EQ5}
\end{equation}
the trace of transfer matrices $|\chi|\leq2$ and corresponding states are extended or critical, where $M$ is the size difference of the two patches, and $\kappa=1,2,\cdots, M-1$ (seen  Appendix B). For a string with more than two patches, using a numerically accurate renormalization scheme~\cite{FA92}, both the sites in intermediate patches and intermediate inlaid sites can be renormalized into ``one'' inlaid site, so they can be taken as ``two patches''. The expression of $E^\kappa_M$ in Eq.(\ref{EQ5}) also holds but $M$ is the size difference of the patches at two edges. This is a local heuristic argument. The details are given in Appendix B. However, for a string with many patches, the renormalized ``one'' inlaid site may induce localization effects.

For a few of patches (super-patches), there are states with energies $E^\kappa_M$. Locally, these states are extended in a few of patches; globally, they are localized in different spaces of the whole lattices. So they are locally-extended localized states. In our model (Fig.~\ref{Fig1}), these patches linearly vary with variation rate $d$. Such locally-extended localized states with same energies $E^\kappa_M$ exist for each super-patches, i.e., resonance conditions, where $M=md$ and $m\in \mathcal{N_{+}}$. When these super-patches are linked together by inlaid B-type sites, the energies of whole lattices around $E^\kappa_M$ may become resonance levels if they are allowed energies, and related states may be extended.

\subsection{Energy spectrum properties}\label{Sec33}

\begin{figure}[!htbp]%fig3
\centering
  \includegraphics[width=1.6in]{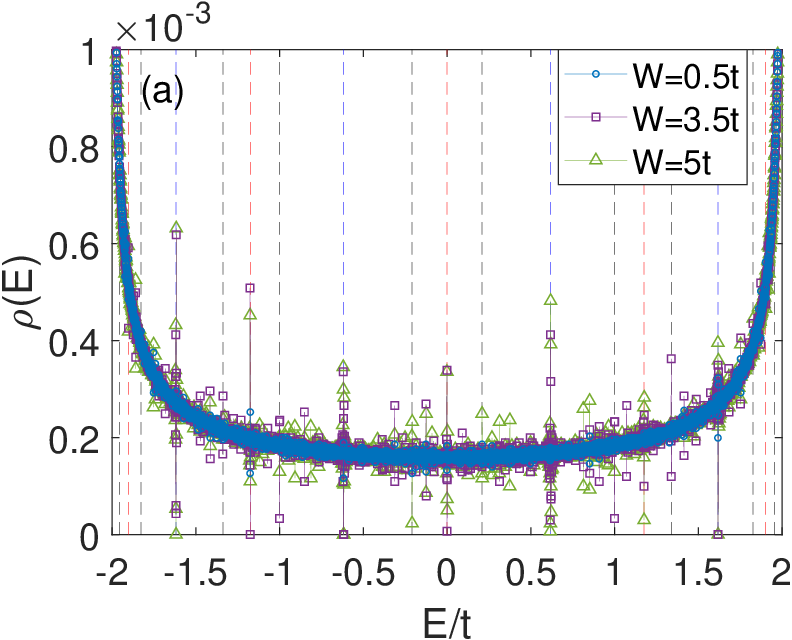}
  \vspace{0.2cm}

  \includegraphics[width=1.6in]{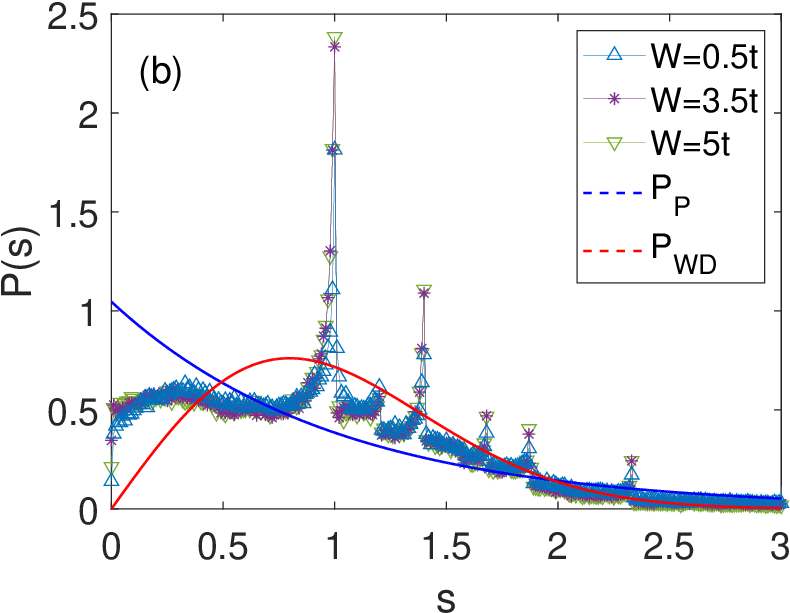}
  ~~~\includegraphics[width=1.6in]{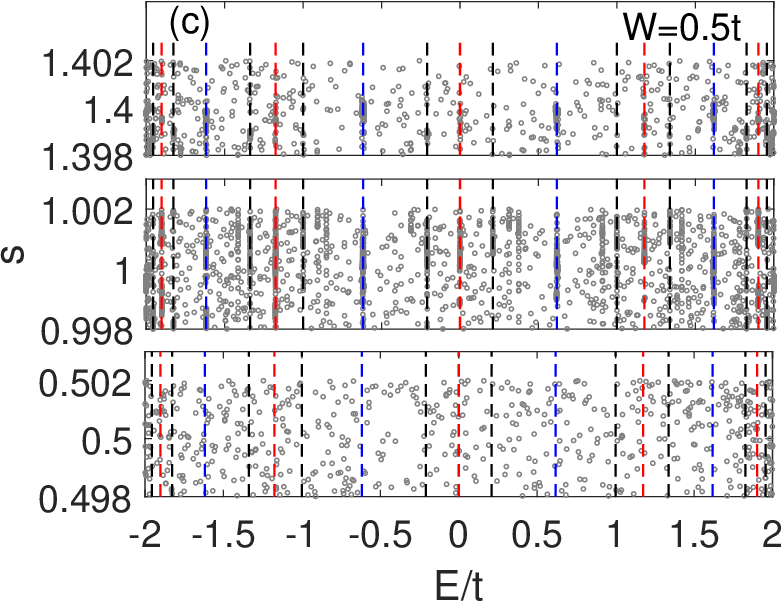}
 \caption{(a) Density of state versus energies $E$ and (b) level spacing distributions $P(s)$, where $W=0.5t, 3.5t$ and $5t$, and system size $N=301,197$ ($j_m=347$). (c) level spacings $s$ versus $E$ at $W=0.5t$. The vertical dashed lines in (a) and (c) are for energies $E^\kappa_M$ with $M=5$ (blue), $M=10$ (red) and $M=15$ (black), respectively. In (b), $P_P$ and $P_{WD}$ are for Poisson and Wigner-Dyson distributions, respectively.}\label{Fig3}
\end{figure}

Statistical properties of energy spectrum can reflect overall properties of systems~\cite{EV08,MI00,CA95,BR81}. Two quantities are of special interest, i.e., the density of state (DOS) and level spacing distributions.

The singularity of DOS can reflect mobility edges in quasiperiodic systems~\cite{DA88} and resonance energies in quantum percolation models~\cite{AL21}.  In Fig.~\ref{Fig3}(a), we plot the DOS $\rho(E)$, which is defined by $\rho(E)=\sum_{\beta=1}^{N}\delta(E-E_\beta)$ and $E_\beta$ is the $\beta$th eigenenergy. We only consider energies $|E|\leq2$. When $E$ is beyond this range, we find states are localized. Fig.~\ref{Fig3}(a) shows in the curves of the $\rho(E)$, there are many sharp peaks along with sharp dips, i.e., the singularity in DOS. The sharp dips mean there may exist energy gaps. Interestingly, most of these peaks (dips) in DOS present at energies that are around $E^\kappa_M$ with $M=5, 10$ and $15$, which are labeled out by vertical dashed lines. The three $M$s are the size difference between nearest-neighbour (NN), next NN, and next to next NN patches. In the figure, the $\kappa=1, 2, 3, 4$ for $M=5$, $\kappa=1, 3, 5, 7, 9$ for $M=10$, and $\kappa=1, 2, 4, 5, 7, 8, 10, 11, 13, 14$  for $M=15$. This can ensure the repeated energies are taken into account only once. For aperiodic systems, such singularities may indicate there exist critical or extended states~\cite{DA88}.

At the same time, for 3D Anderson models, level spacing distributions $P(s)$ are the Wigner-Dyson distribution $P_{WD}$ in a metal region, obey Poisson law $P_P$ in an insulator region, and they are intermediate at the metal-insulator transitions~\cite{SH93}, where $s$ are spacings between unfolded nearest neighboring levels. The $P(s)$  are also intermediate distributions in disorder systems with long-range correlations~\cite{CA04} and in quasiperiodic systems such as Fibonacci and Thue-Morse chains~\cite{CA95}. We plot $P(s)$ in Fig.~\ref{Fig3}(b) for $W=0.5t$, $3.5t$ and $5t$, respectively. It shows the behaviors of $P(s)$ are intermediate between $P_{WD}$ and $P_P$. Such behaviors are consistent with that the structures in Fig.~\ref{Fig1} are aperiodic. Further, Fig.~\ref{Fig3}(b) shows $P(s)$ have local sharp peaks at some of $s$. In Fig.~\ref{Fig3}(c), we show the two main peaks at $s=1$ and $1.4$ are mainly attributed to these energies with singularity in DOS. As a comparison, at $s=0.5$ [without peaks in $P(s)$], their attributions are common.

\subsection{State localization properties}\label{Sec34}
We use three effective quantities, i.e., the local tensions~\cite{DE19,EV22}, Lypanunov exponents~\cite{FA92} and fractional dimensions~\cite{EV08}, to directly characterize state localization properties.

\subsubsection{Local tension}\label{Sec341}
Firstly, the local tension successfully distinguishes metals from insulators in many-body systems~\cite{DE19}. For 1D systems, it is defined by $\lambda_{xx}^2=\langle\Psi|\hat{Q}_x^{\dag}\hat{Q}_x|\Psi\rangle-\langle\Psi|\hat{Q}_x^{\dag}|\Psi\rangle\langle\Psi|\hat{Q}_x|\Psi\rangle$, where $\hat{Q}_x=\frac{N}{2{\pi}i}\sum_{n=0}^{N-1}[\exp(\frac{2{\pi}i}{N}\hat{x}_n)-1]$,
and $\hat{x}_n$ is the position operator in the x-direction~\cite{CH23}, i.e., $\hat{x}_n=|n \rangle$. The reduced local tension (RLT) is defined by
\begin{equation}
\Lambda=2\pi\lambda_{xx}/N\label{EQ6}.
\end{equation}
As system sizes $N\to\infty$, $\Lambda\to1$ for extended states and $\Lambda\to0$ for localized ones~\cite{TA22}.

%----------------
\begin{figure}[!htbp]%fig4
\centering
\includegraphics[width=1.68in]{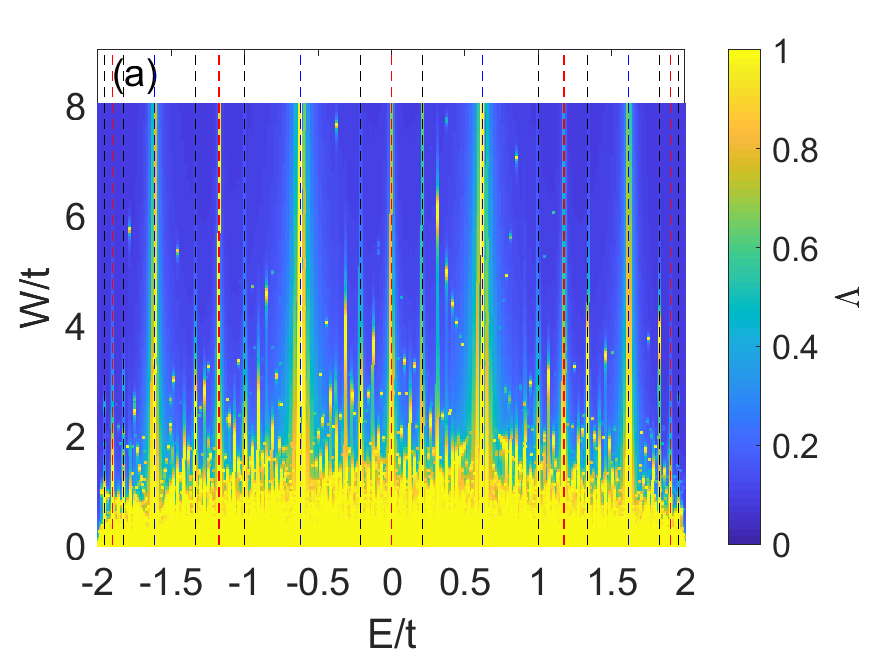}
\includegraphics[width=1.58in]{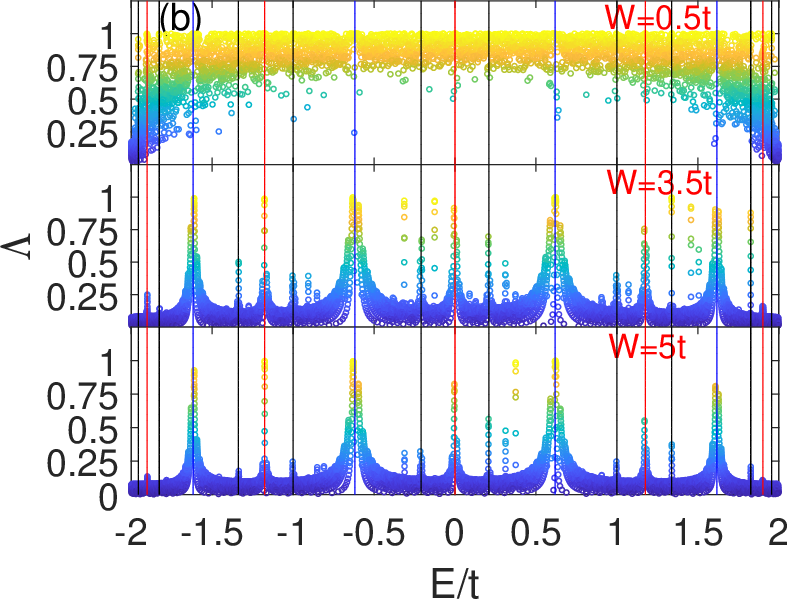}
\vspace{0.2cm}

\includegraphics[width=1.5in]{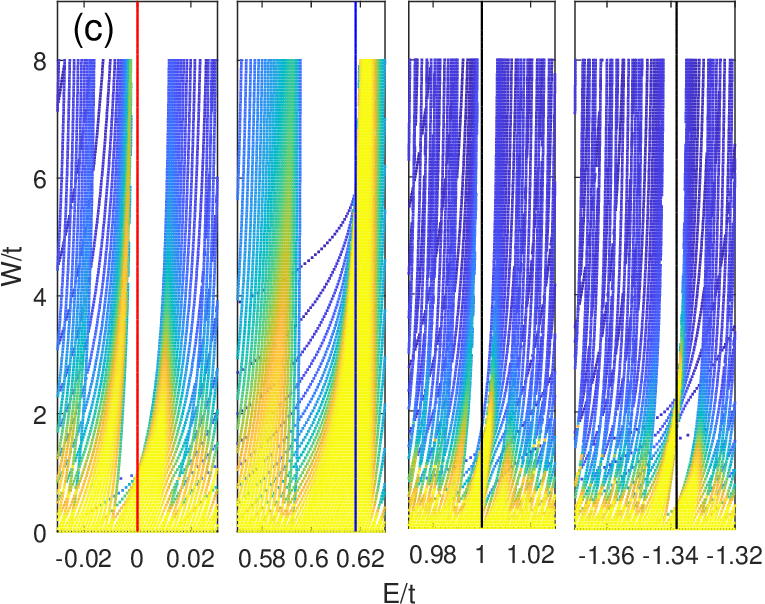}
\includegraphics[width=1.65in]{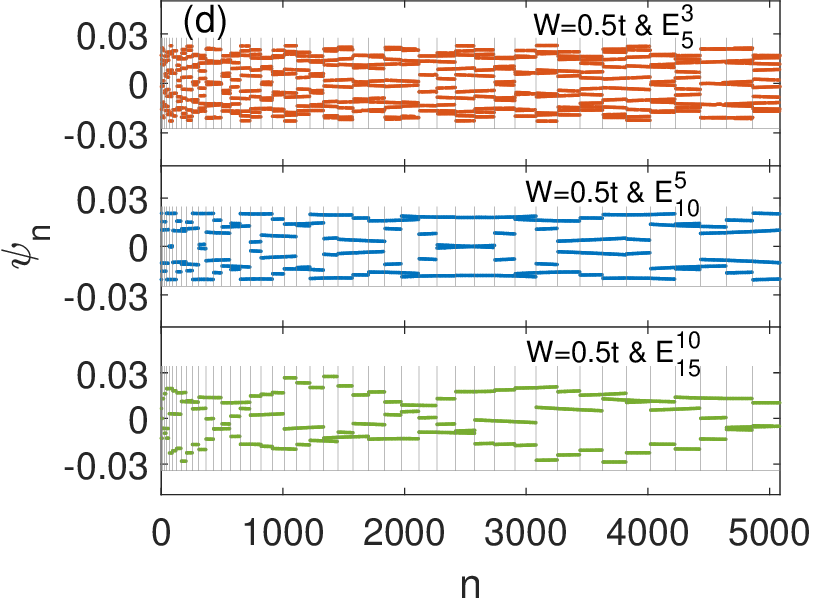}
 \caption{(a) RLTs $\Lambda$ as functions of energies $E$ and potential strengths $W$. (b) The $\Lambda$ versus $E$ at $W=0.5t, 3.5t$ and $5t$, respectively. (c) From left to right, partial enlarger for $E$ near $E^5_{10}=0$, $E^3_5\approx0.618t$, $E^{10}_{15}=t$ and $E^4_{15}\approx-1.338t$, respectively. (d) At $W=0.5t$, three typical wave functions with eigenenergies nearest to $E^3_5$, $E^5_{10}$ and $E^{10}_{15}$, respectively. The vertical dashed lines in (a)-(c) are the same in Fig.~\ref{Fig3}. The grey fold lines in (d) are for the functions of on-site potentials. System size $N=5,086$ ($j_m=45$).}\label{Fig4}
\end{figure}

Fig.~\ref{Fig4}(a) shows generally, the $\Lambda$ are relatively large (small) for relative small (large) $W$. The vertical dashed lines mark the position of $E^\kappa_M$ obtained from Eq.(\ref{EQ5}). The $\Lambda$ are relatively large when $E$ are around these $E^\kappa_M$ [also seen Fig.~\ref{Fig4}(b)], which means these states may be extended. Partial enlargers of Fig.~\ref{Fig4}(a) for $E$ near four $E^\kappa_M$ are plotted in Fig.~\ref{Fig4}(c). It shows when $W$ are relative large, energy gaps along with level squeezing will occur, which corresponds to the singularity in DOS shown in Fig.~\ref{Fig3}(a); for these squeezed levels, the values of $\Lambda$ may be relatively large; in these energy gaps, states with energies $E^\kappa_M$ are not permitted. In Fig.~\ref{Fig4}(d), we plot typical wave functions with $E$ that are nearest to three $E^\kappa_M$, where they spread over the whole lattices, i.e., they are extended states. In fact, we find the same results for other cases, including different $d_0$ at $d=5$ and  other $d$s  (seen Appendix C).

\begin{figure}[!htbp]%fig5
\centering
  \includegraphics[width=1.6in]{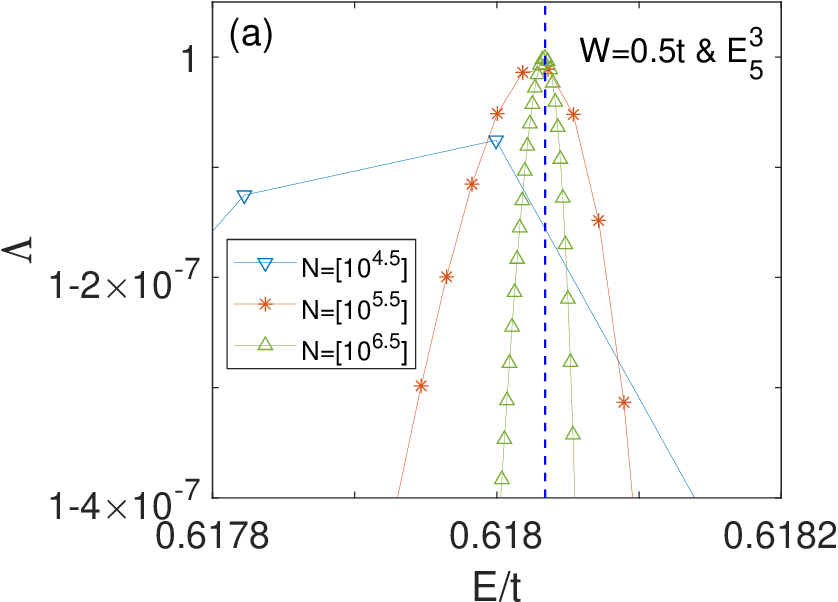}
  \includegraphics[width=1.53in]{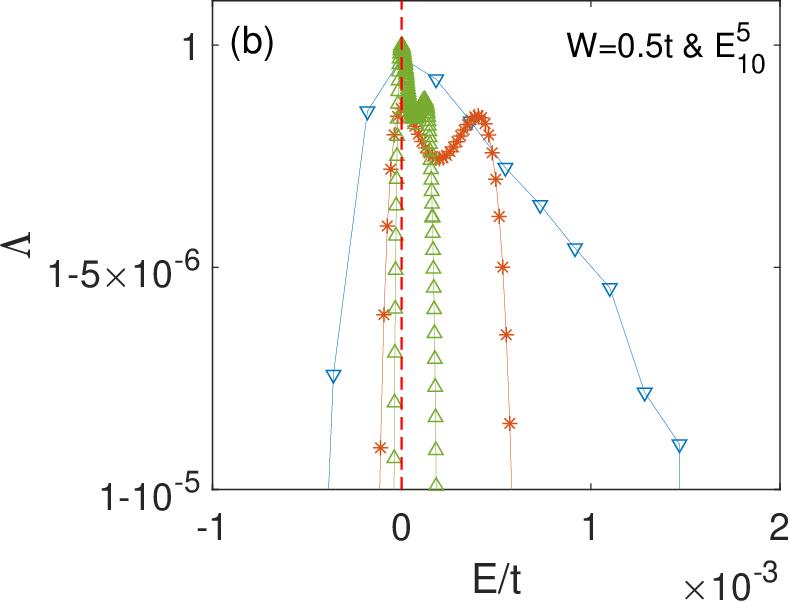}
 \vspace{0.2cm}

 \includegraphics[width=1.6in]{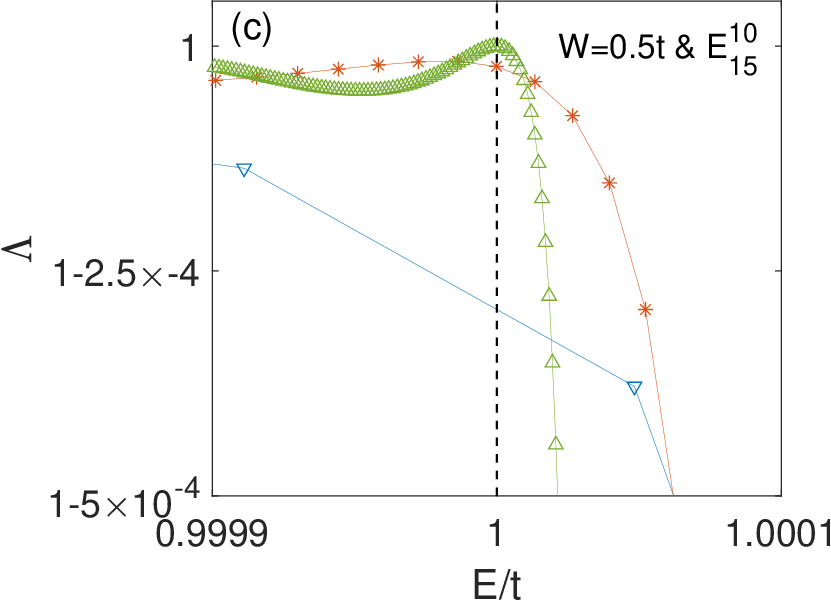}
 \includegraphics[width=1.45in]{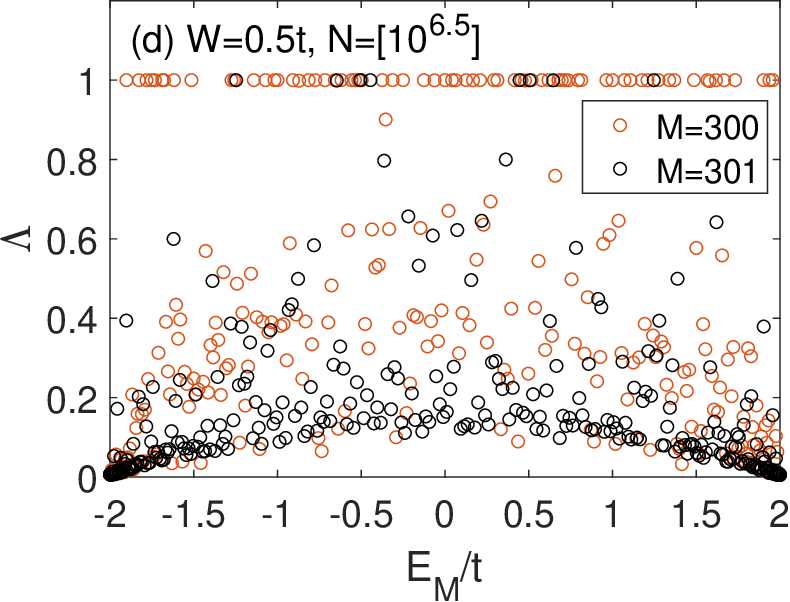}
 \caption{At $W=0.5t$, RLTs $\Lambda$ as functions of energies $E$ when they are near (a) $E^3_5$, (b) $E^5_{10}$ and (c) $E^{10}_{15}$, respectively. (d) The $\Lambda$ versus $E^\kappa_M$ with $M=300$ and $301$. The vertical dashed lines in (a)-(c) are the same in Fig.~\ref{Fig3}. System sizes $N=[10^{4.5}]=31,980~(j_m=227), [10^{5.5}]=317,019~(j_m=356)$  and $[10^{6.5}]=3,164,626~(j_m=1125)$. Here, $[10^{z}]$ denotes the system size that is greater than and nearest to $10^{z}$.}
 \label{Fig5}
\end{figure}

\begin{figure}[!htbp]%fig6
\centering
  \includegraphics[width=1.6in]{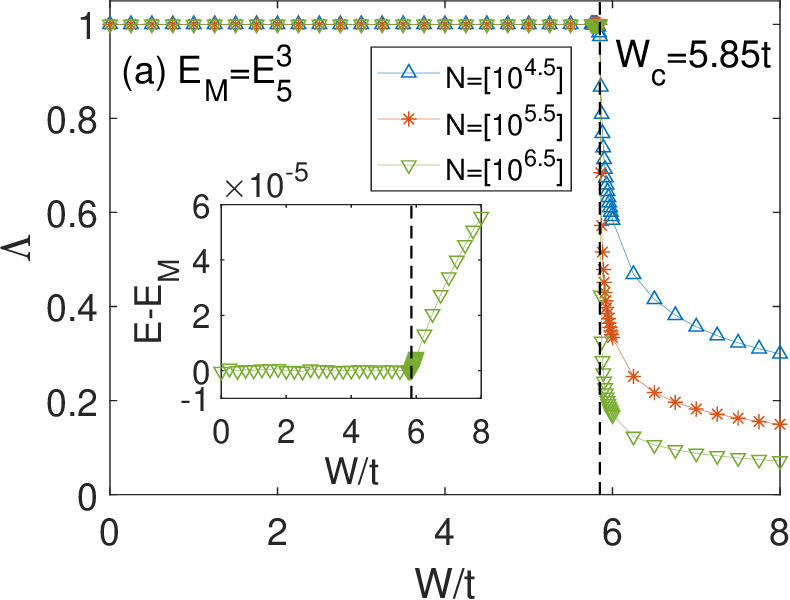}
  \includegraphics[width=1.6in]{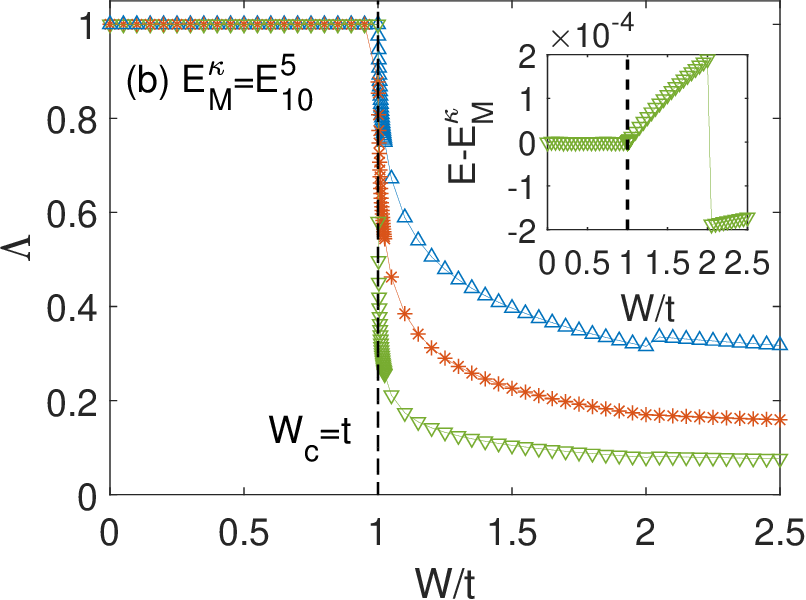}
  \vspace{0.2cm}

  \includegraphics[width=1.6in]{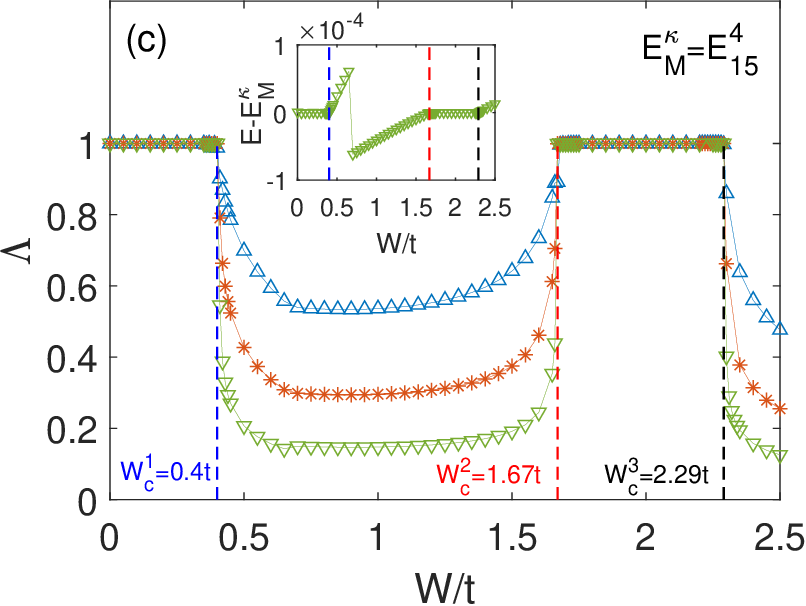}
  \includegraphics[width=1.6in]{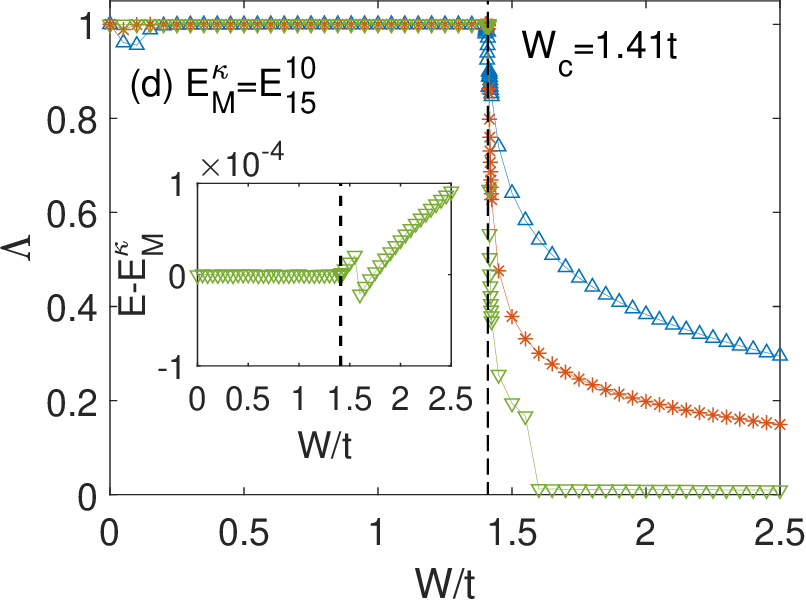}

  \vspace{0.2cm}
  \includegraphics[width=1.6in]{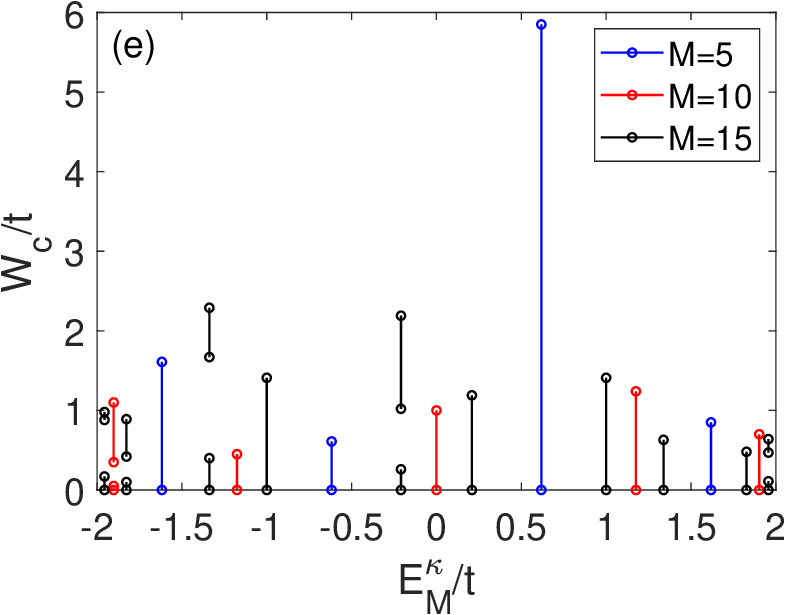}

 \caption{RLTs $\Lambda$ as functions of potential strengths $W$ at energies $E$ that are nearest to (a) $E^3_5$, (b) $E^5_{10}$, (c) $E^{4}_{15}$ and (d) $E^{10}_{15}$, respectively. (e) $W_c$ versus $E^\kappa_M$ at $M=5, 10$ and $15$, and lines represent the range of $W$ where states are extended. The insets in (a)-(d) show the $E-E^\kappa_M$ versus $W$ at $N=[10^{6.5}]$. The vertical dashed lines in (a)-(d) mark the position of $W_c$.}\label{Fig6}
\end{figure}

Figs.~\ref{Fig5}(a)-(c) show at $W=0.5t$, as system sizes $N$ increase, the values of $\Lambda$ are close to ones for states at energies $E^\kappa_M=E^3_5, E^5_{10}$ and $E^{10}_{15}$, respectively, while they decrease when states with energies $E$ depart from these $E^\kappa_M$. In Fig.~\ref{Fig5}(d), we plot the $\Lambda$ versus $E^\kappa_M$ with $M=300$. In calculations, these states are the ones with energies nearest to $E^\kappa_M$. It shows most of $\Lambda$ almost equal to ones, which correspond to extended states; many are much smaller than ones, which correspond to localized states. As a comparison, we also plot the $\Lambda$ at $M=301$. The value $301$ is not the size difference of patches, so the corresponding $E^\kappa_M$ are not resonance energies. For them, Fig.~\ref{Fig5}(d) shows almost all $\Lambda$ are smaller than ones, which means these states are localized.

In Fig.~\ref{Fig6}, we plot the $\Lambda$ as functions of $W$ when $E$ are nearest to $E^\kappa_M$. It shows the values of $\Lambda$ are close to ones in some regions of $W$, where states are extended. We call them  ``extended-state-W regions''. And $\Lambda$ rapidly decrease at the boundaries of such regions. These are the signatures of delocalization-localization transitions, and the corresponding critical potential strength is denoted by $W_c$. So the states have delocalization-localization transitions at $W_c$. We plot the $E-E_M$ versus $W$ in the insets of Figs.~\ref{Fig6}(a)-(d) at $N=[10^{6.5}]$. We find in extended-state-W regions, $E-E^\kappa_M\approx0$, i.e. $E^\kappa_M$ is allowed energies of systems. In localized state regions of $W$, $E-E^\kappa_M$ are relative large; as displayed in Fig.~\ref{Fig4}(c), these $E^\kappa_M$ are in energy gaps. We plot the phase diagram in Fig.~\ref{Fig6}(e) for $M=5, 10$ and $15$, where lines represent the ranges of extended-state-W regions. Similarly, we can obtain phase diagrams for larger $M$, but the corresponding ranges are relative small, or even disappear.

\subsubsection{Lyapunov exponent}\label{Sec342}

Secondly, the energy-dependent Lyapunov exponent (LE) $\gamma(E)$ is another often used quantity to characterize electronic localization properties, which is defined by
\begin{equation}
\gamma(E)=-\lim_{N\to\infty}\bigg[\frac{1}{N}\ln\Big|\frac{G_{NN}(E)}{G_{1N}(E)}\Big|\bigg], \label{EQ7}
\end{equation}
where $G_{NN}(E)$ and $G_{1N}(E)$ are the Green-function matrix elements. We use a numerically accurate renormalization scheme to calculate them~\cite{FA92}. Generally, the LE $\gamma$ is inversely proportional to localization length. At finite system sizes for extended states, $\gamma$ may be less than zeros. In practice, for finite system size the inequation $\gamma\leq1/N$ is often used as a sign that states are extended.

\begin{figure}[!htbp]%fig7
\centering
 \includegraphics[width=1.6in]{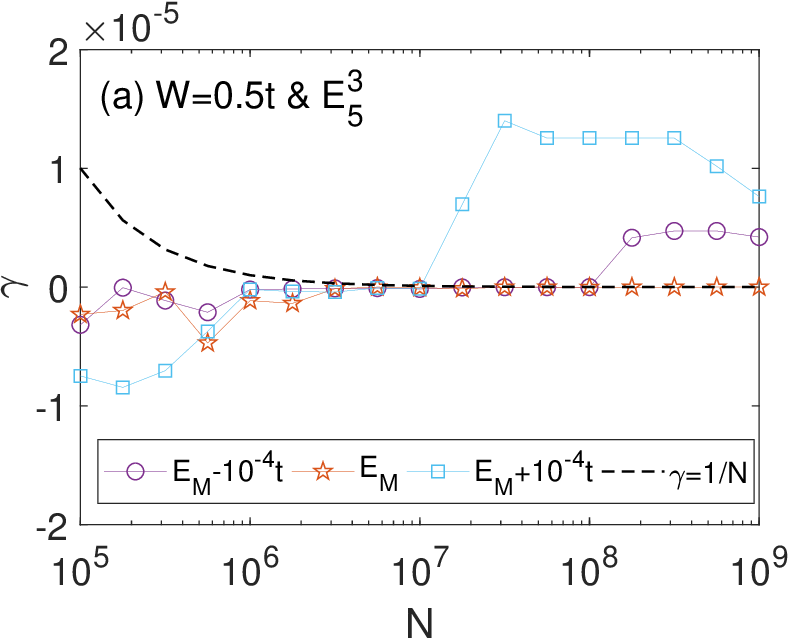}
 \includegraphics[width=1.6in]{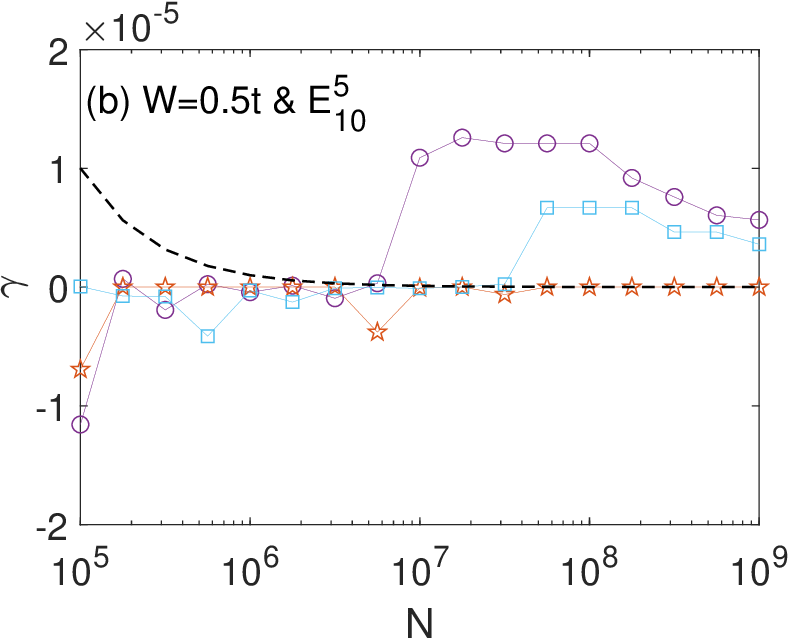}
 \vspace{0.2cm}

 \includegraphics[width=1.6in]{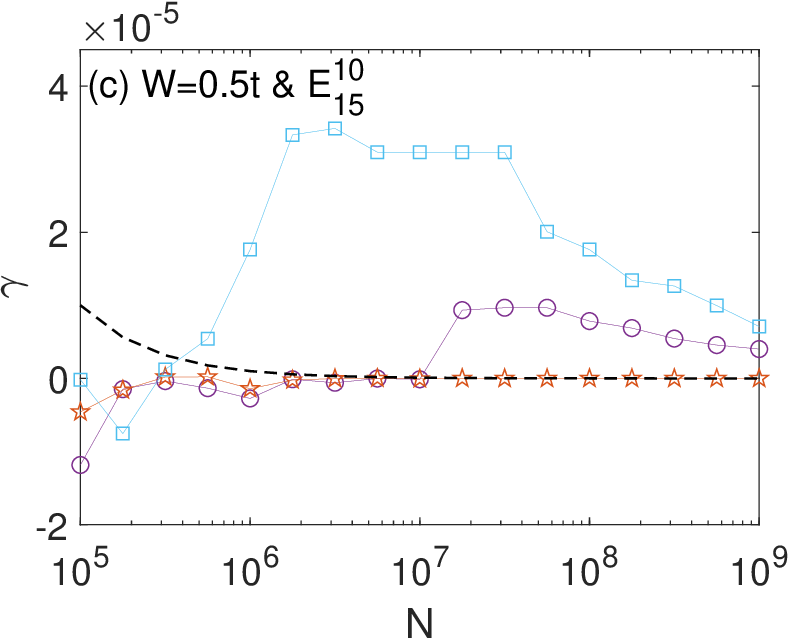}
 \includegraphics[width=1.6in]{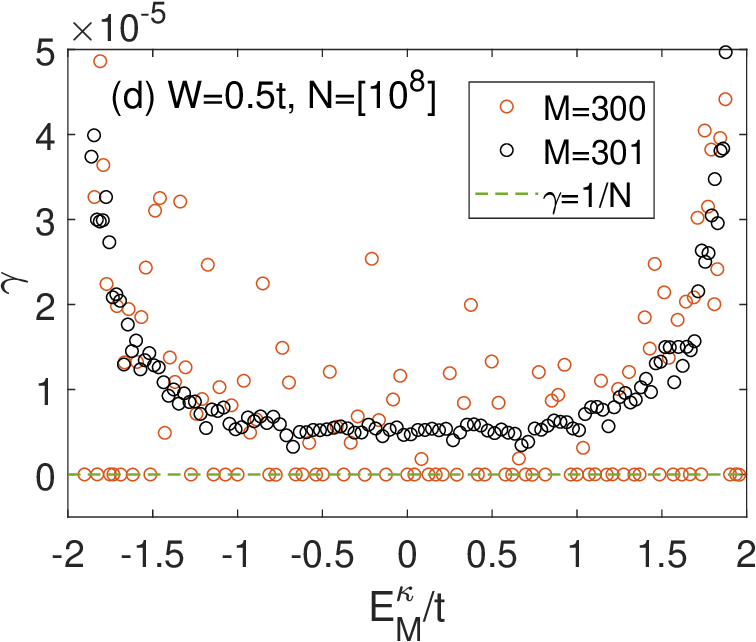}
 \caption{At $W=0.5t$, LEs $\gamma$ versus system sizes $N$ with energies $E$ at and near (a) $E^3_5$, (b) $E^5_{10}$ and (c) $E^{10}_{15}$, respectively. The $\gamma$ versus $E^\kappa_M$ with $M=300$ and $301$. The dashed curves in (a)-(d) are for the function $\gamma=1/N$. }\label{Fig7}
\end{figure}
%%%%%%%%%%%%%
\begin{figure}[!htbp]%fig8
\centering
\includegraphics[width=1.6in]{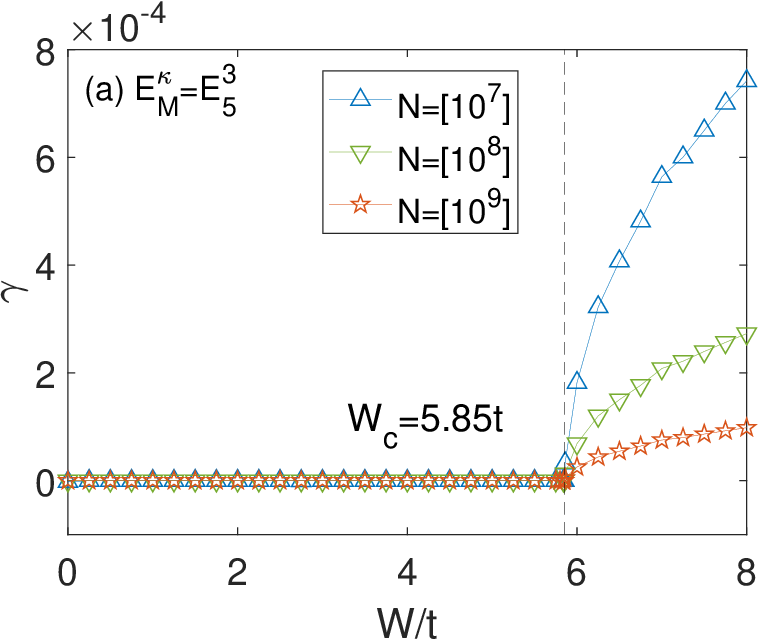}
\includegraphics[width=1.6in]{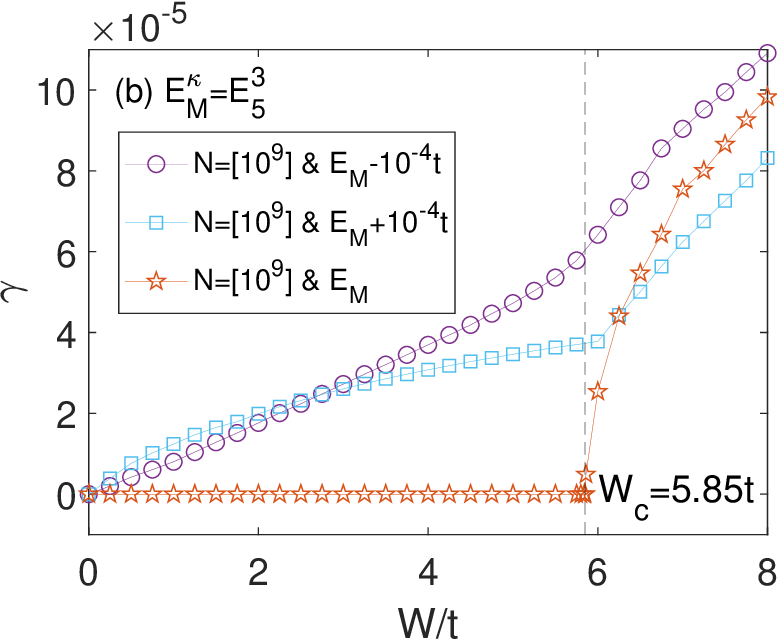}
\vspace{0.2cm}

\includegraphics[width=1.6in]{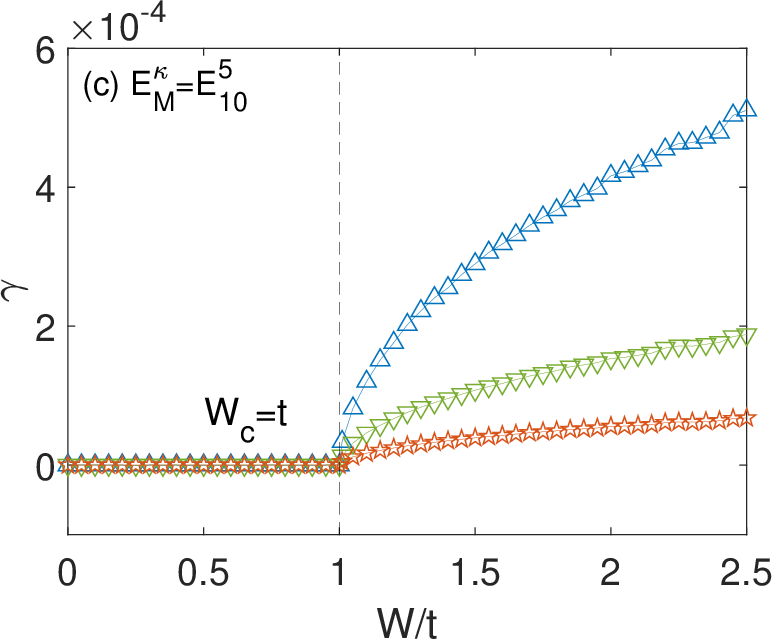}
\includegraphics[width=1.6in]{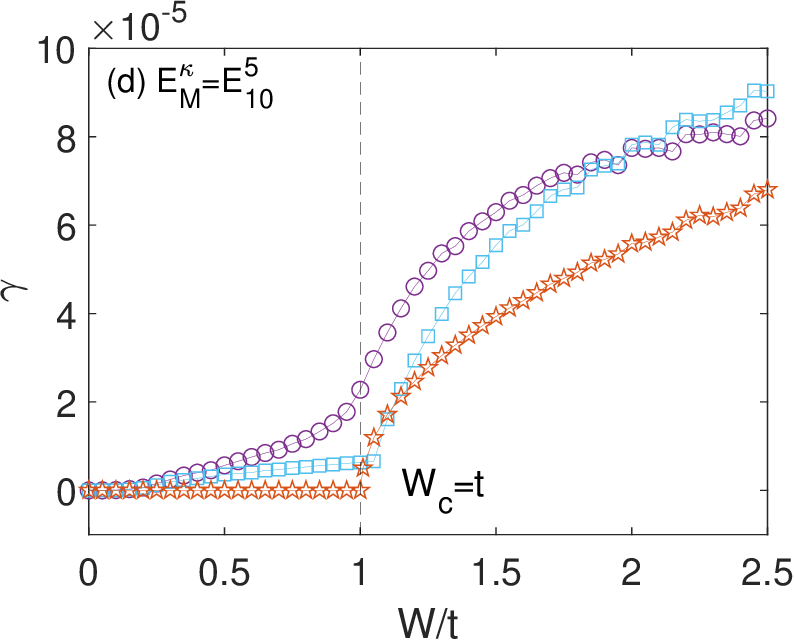}
\vspace{0.2cm}

\includegraphics[width=1.6in]{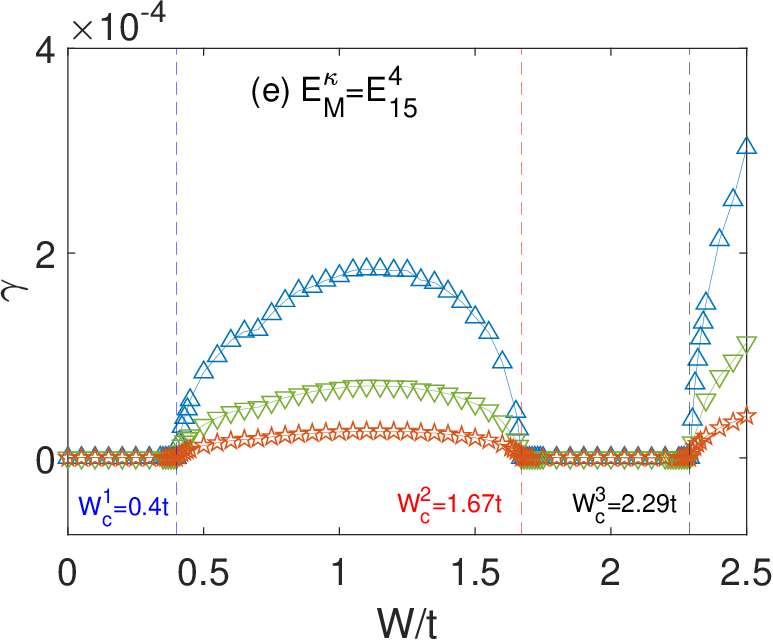}
\includegraphics[width=1.6in]{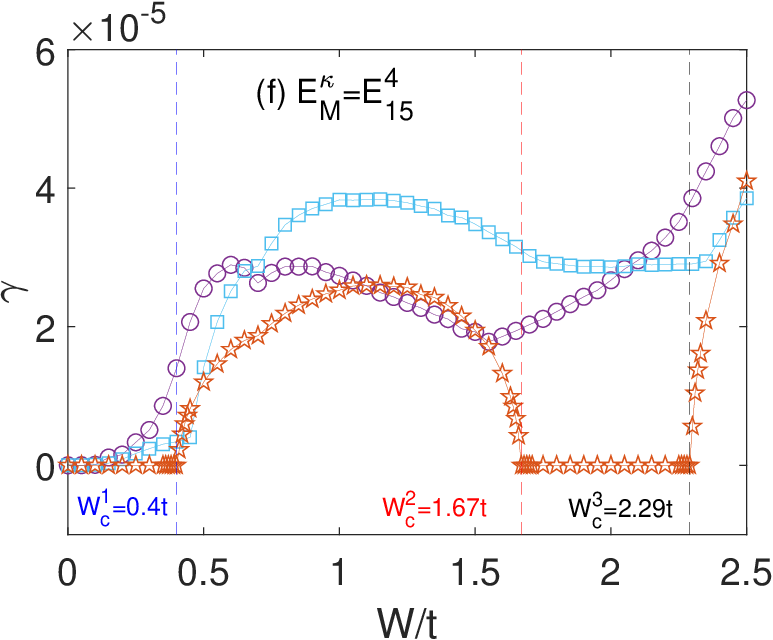}
\vspace{0.2cm}

\includegraphics[width=1.6in]{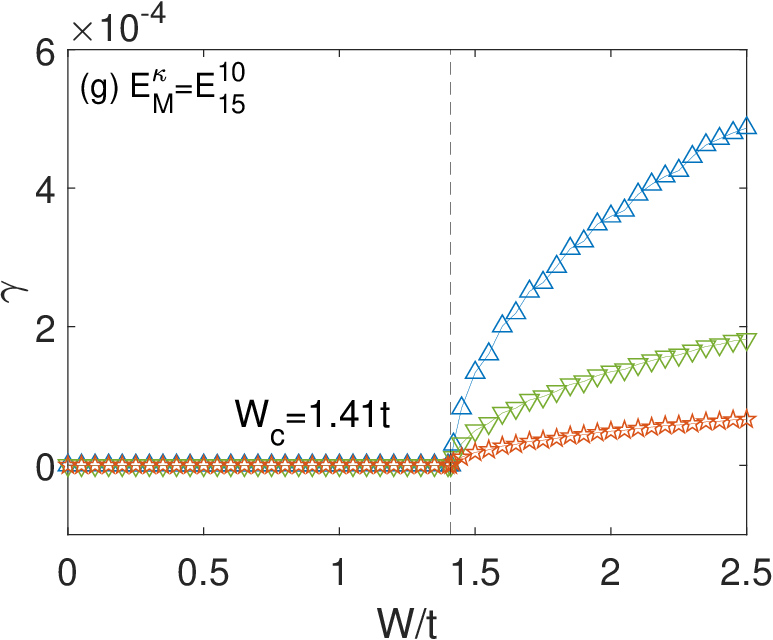}
\includegraphics[width=1.6in]{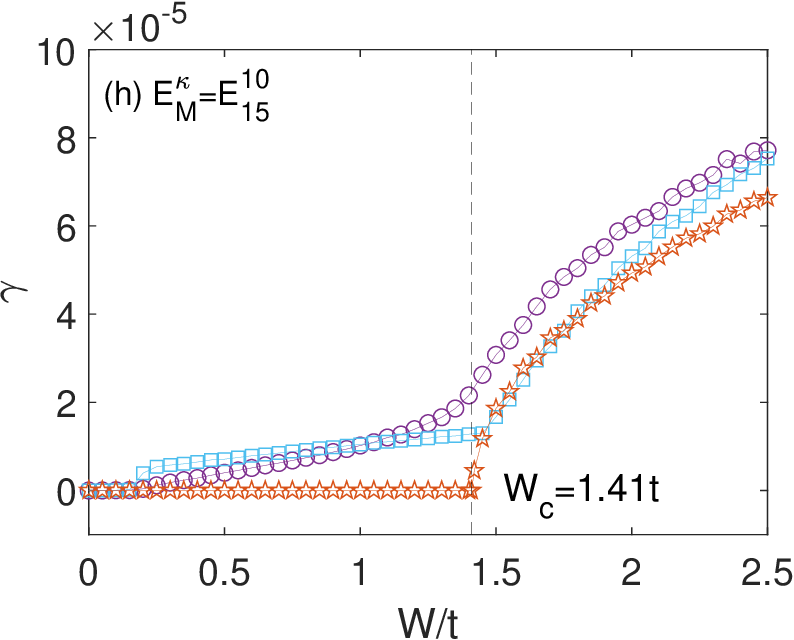}
\caption{LEs $\gamma$ as functions of potential strength $W$ for energies $E$ at and near (a, b) $E^3_5$, (c, d) $E^5_{10}$, (e, f) $E^{4}_{15}$ and (g, h) $E^{10}_{15}$, respectively. The vertical dashed lines in (a)-(d) mark the position of $W_c$.}\label{Fig8}
\end{figure}
%%%%%%%%%%%%%
\begin{figure}[!htbp]%fig9
\centering
 \includegraphics[width=1.6in]{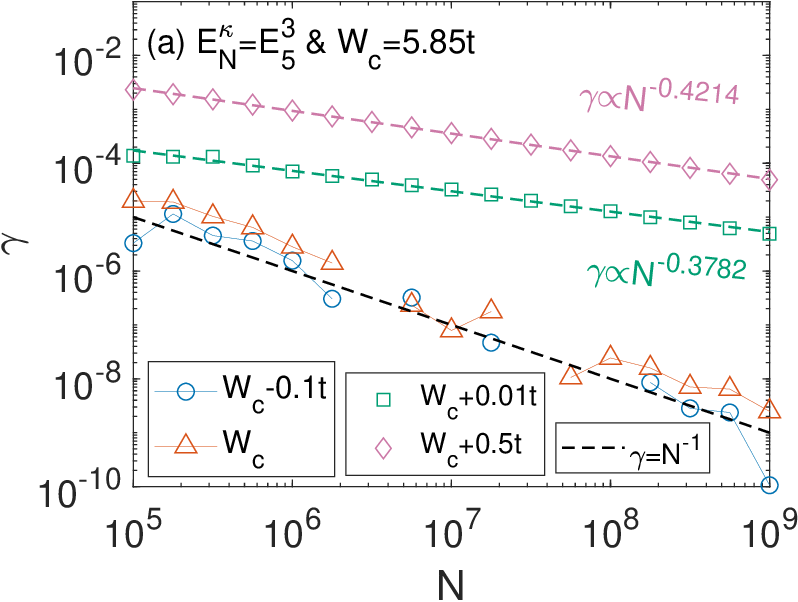}
 \includegraphics[width=1.6in]{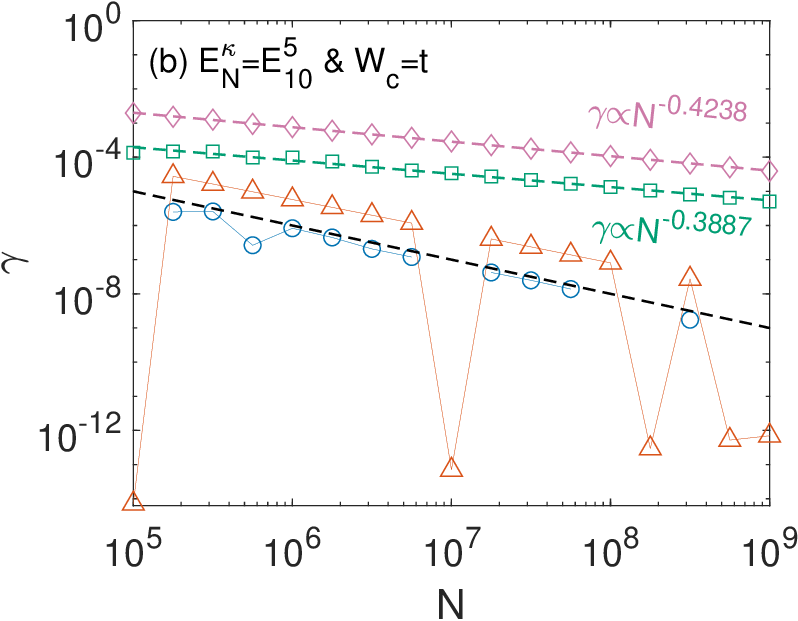}
 \vspace{0.2cm}

 \includegraphics[width=1.6in]{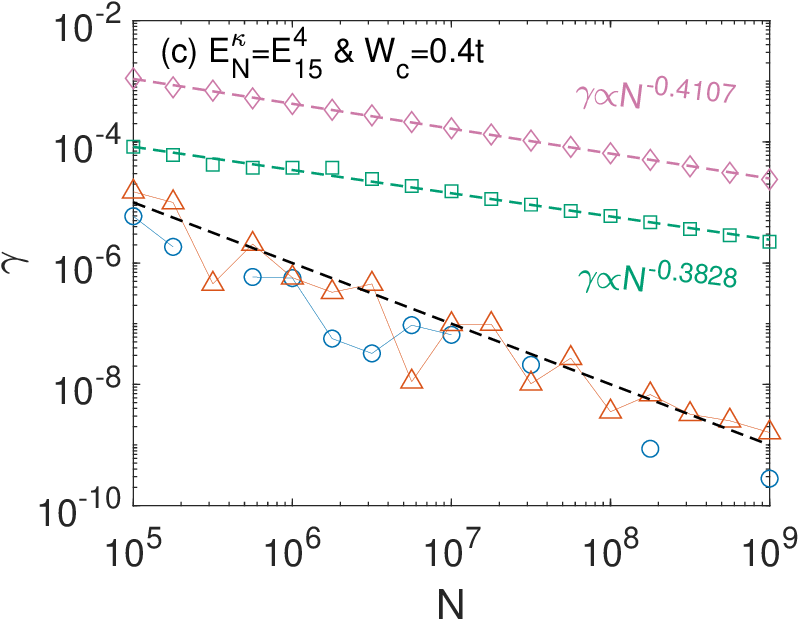}
 \includegraphics[width=1.6in]{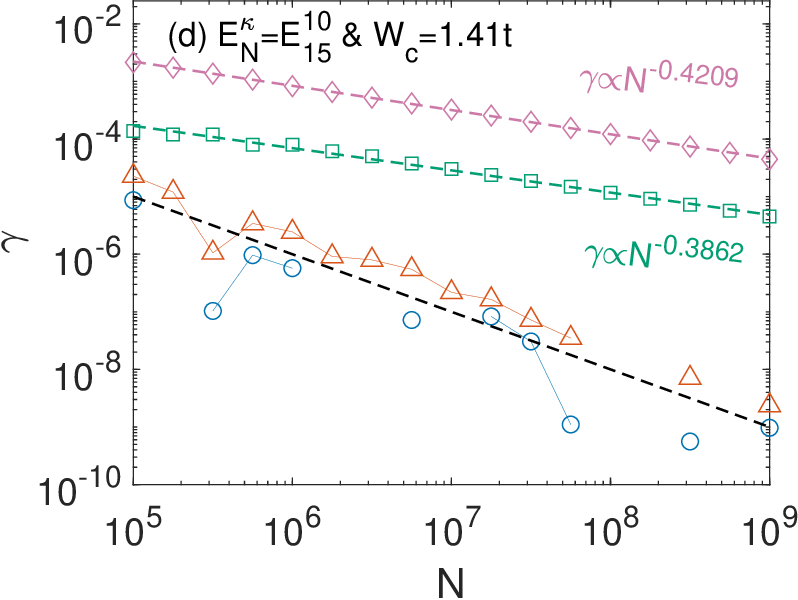}
 \caption{LEs $\gamma$ versus system sizes $N$ with potential strength at critical $W_c$, $W_c-0.1t$, $W_c+0.01t$ and $W_c+0.5t$, respectively. Energies $E$ are at (a) $E^3_5$, (b) $E^5_{10}$, (c) $E^{4}_{15}$ and (d) $E^{10}_{15}$, respectively, where the dashed lines are for the function $\gamma\propto N^{-\nu}$.
 }\label{Fig9}
\end{figure}
%%%%%%%%%%%%%
%%%%%%%%%%%%%
\begin{figure}[!htbp]%fig10
\centering
 \includegraphics[width=1.6in]{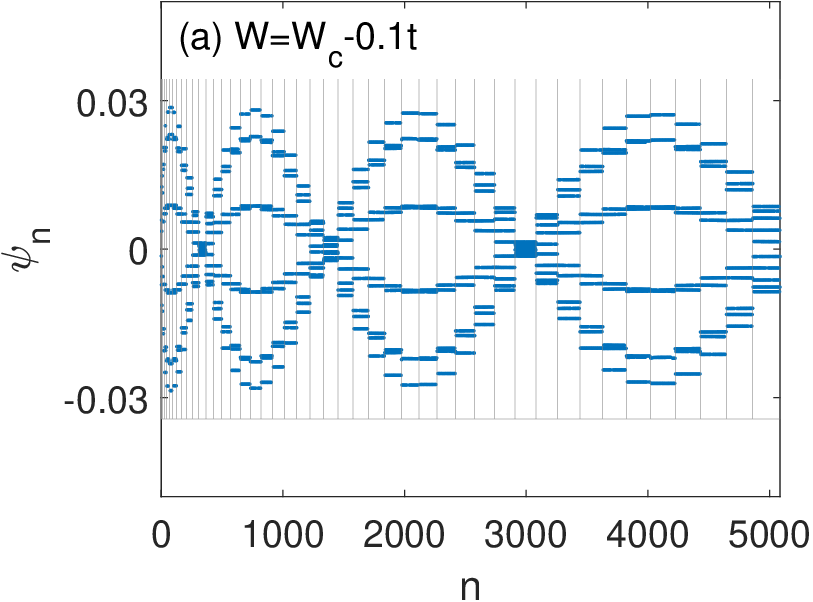}
 \includegraphics[width=1.6in]{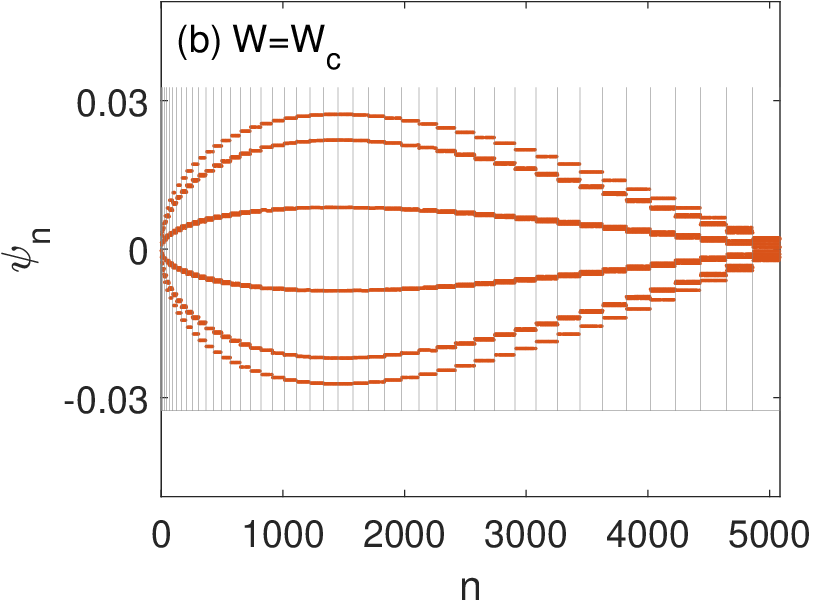}
 \vspace{0.2cm}

 \includegraphics[width=1.6in]{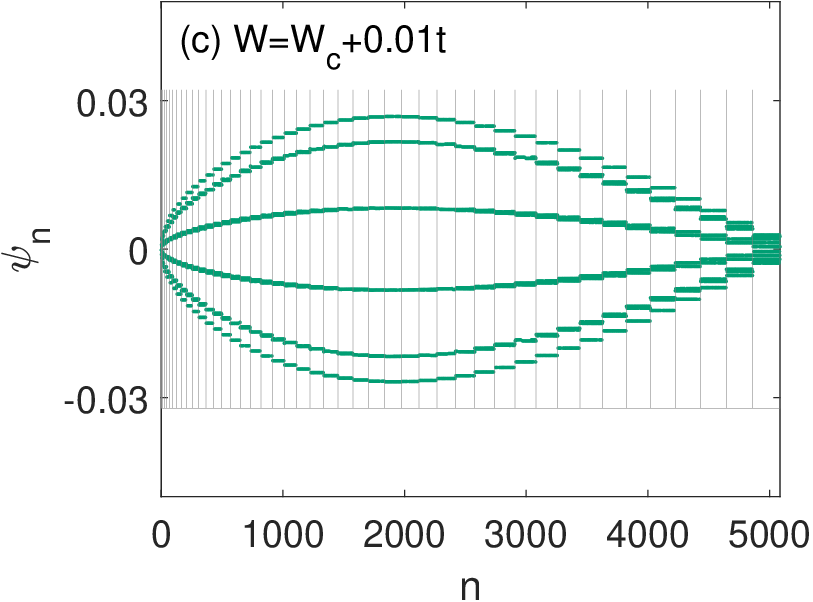}
 \includegraphics[width=1.6in]{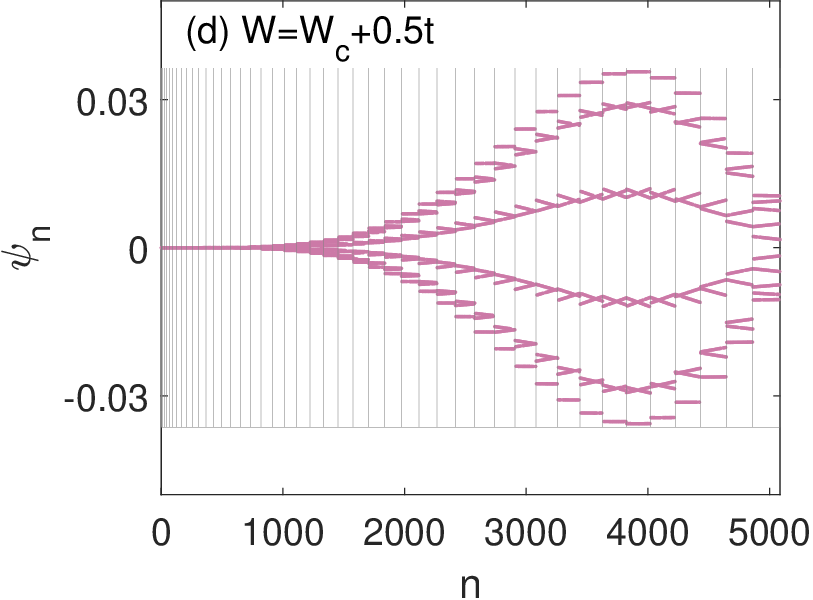}
 \vspace{0.2cm}

 \includegraphics[width=1.6in]{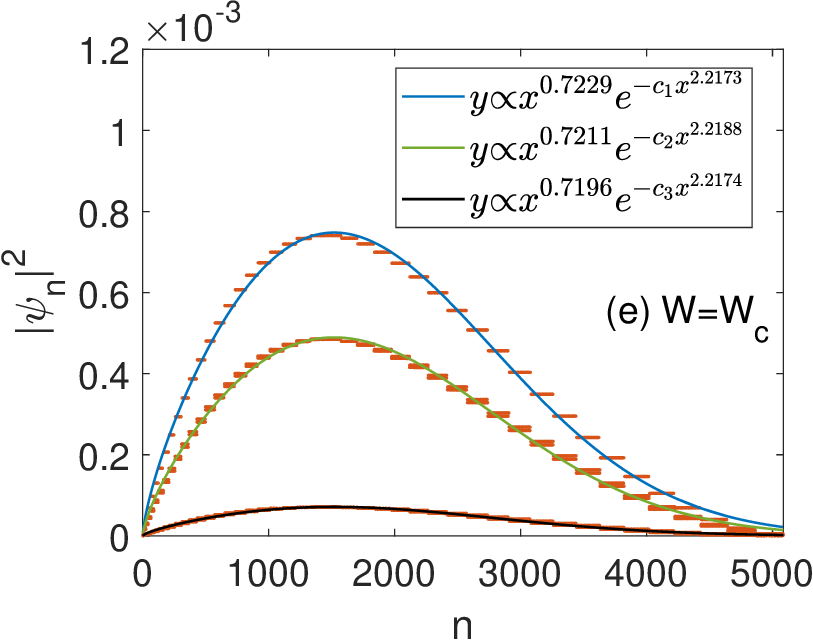}
 \includegraphics[width=1.6in]{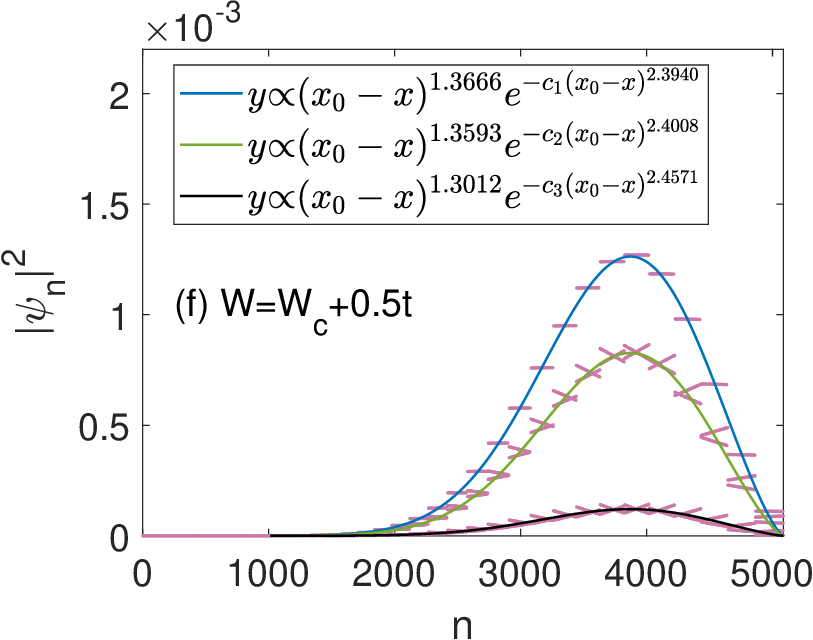}
 \caption{Typical wave functions with eigenenergies nearest to $E^3_5$ when $W$ equals to (a) $W_c-0.1t$, (b) $W_c$, (c) $W_c+0.01t$ and (d) $W_c+0.5t$, respectively, where $W_c=5.85t$. The square moduli of wave functions when (e) $W=W_c$ and (f) $W=W_c+0.5t$. The grey fold lines in (a)-(d) are for the functions of on-site potentials. In (e), $c_1=2.8712\times10^{-8}$, $c_2=2.8321\times10^{-8}$ and $c_3=2.8578\times10^{-8}$. In (f), $c_1=2.3522\times10^{-8}$, $c_2=2.2195\times10^{-8}$, $1.3897\times10^{-8}$ and $x_0=5087$. System size $N=5,086$ ($j_m=45$). System size $N=5,086$ ($j_m=45$).}\label{Fig10}
\end{figure}
%%%%%%%%%%%%%

Figs.~\ref{Fig7}(a)-(c) show $\gamma\leq1/N$ when energies at $E^\kappa_M$; as system sizes $N$ are large enough, $\gamma>1/N$ when energies deviate from $E^\kappa_M$ even a little bit. This implies $E^\kappa_M$ are discrete resonance levels. In Fig.~\ref{Fig7}(d), we plot the $\gamma$ versus $E^\kappa_M$ with $M=300$. It shows most of $\gamma$ are smaller than $1/N$, which indicates these states are extended. At the same time, many are larger than $1/N$. We also plot the $\gamma$ at $M=301$. Conversely, all values of $\gamma$ are larger than $1/N$, i.e., all these states are localized ones.

Figs.~\ref{Fig8}(a), (c), (e) and (g) show as $N$ increases, there exist regions that $\gamma\to0$ and $\gamma$ are finite. The $W_c$ separates the two types of regions, which agrees with that shown in Fig.~\ref{Fig6}. At the same time, Figs.~\ref{Fig8}(b), (d), (f) and (h) plot $\gamma$ versus $W$ at $E^\kappa_M$ and $E^\kappa_M\pm10^{-4}t$. They show in extended-state-W regions, when energies deviate from $E^\kappa_M$ a little bit, $\gamma$ are finite, which means these states are localized.

Figs.~\ref{Fig9}(a)-(d) show the behaviour of the $\gamma$ with respect to system sizes $N$ when potential strength at $W_c$, $W_c-0.1t$, $W_c+0.01t$ and \textcolor{blue}{$W_c+0.5t$}, respectively. As the logarithm is applied, the $\gamma$ for some $N$ are not displayed if their values are smaller than zeros. Theses figures show when $W$ are in the extended-state-W regions, generally, $\gamma{\propto}N^{-1}$, which confirms these states are extended. At the same time, when $W$ are beyond such regions, $\gamma{\propto}N^{-\nu}$ and scaling exponents $\nu$ are less than $1$, which indicates these states are localized. To demonstrate the localization properties intuitively, we plot typical wave functions with eigenenergies nearest to $E^3_5$ in Fig.~\ref{Fig10}. We find the state in Fig.~\ref{Fig10}(a) is an extended state, which spreads over the whole lattices. The states in Figs.~\ref{Fig10}(b) and (d) are critical (intermediate) and localized ones, respectively. In Fig.~\ref{Fig10}(c), the varying of the state (with $W$ being close to $W_c$) is similar as that in Fig.~\ref{Fig10}(b). From Fig.~\ref{Fig9}(a), we can infer such state should be localized for the corresponding $\nu$ is much less than $1$ when $N$ is large enough. At the same time, for critical and localized states, the square moduli of wave functions are plotted in Figs.~\ref{Fig10}(e) and (f), which exhibit three hierarchies and may indicate such wave functions have fractal properties. Interestingly, every hierarchy can be fitted by the function $y=ax^{b_1}\exp(-cx^{b_2})$. In 1D systems, for exponentially localized states~\cite{KR93}, LEs $\gamma$ shall remain finite when $N\to\infty$, so $\nu=0$; for power-law localized states~\cite{VA92}, $\gamma\to0$ as $N\to\infty$, but $\nu$ are finite. Different from the two cases, in the present work, critical and localized states can be described by the power-law function tuned with exponential decay functions. Lyapunov exponents can also be calculated by~\cite{VA92} $\gamma=\frac{1}{2N}\sum_{x=1}^N{\ln{\frac{y(x)}{y(x-1)}}}$, so it can be written as $\gamma=\gamma_{power}+\gamma_{exp}$. As $N$ increase, the former determines the scaling property of $\gamma$, and the latter determines the upper bound of $\gamma$. This agrees with that shown in Fig.~\ref{Fig9}, where the scaling exponents $\nu$ are close to ones for critical states and they are much less than ones for localized states.

\subsubsection{Fractional dimension}\label{Sec343}

Thirdly, do these resonance states keep extended in the thermodynamic limit? We shed light on this problem with fractional dimension (FD)~\cite{EV08}, which is defined by $D_q=\frac{1}{1-q}\ln{I_q}/\ln(N)$, where
$I_q=\sum_{n=0}^N|\psi_\beta(n)|^{2q}$ and $q$ is the moment. In the thermodynamic limit,
 \begin{equation}
D_q^{\infty}=\lim_{N\to\infty}D_q. \label{EQ8}
\end{equation}
In 1D systems, $D_q^{\infty}=1$ for perfectly extended states, $D_q^{\infty}=0$ for localized states, and $0<D_q^{\infty}<1$ for intermediate ones~\cite{EV08,AH22}.

\begin{figure}[!htbp]%fig11
\centering
 \includegraphics[width=1.6in]{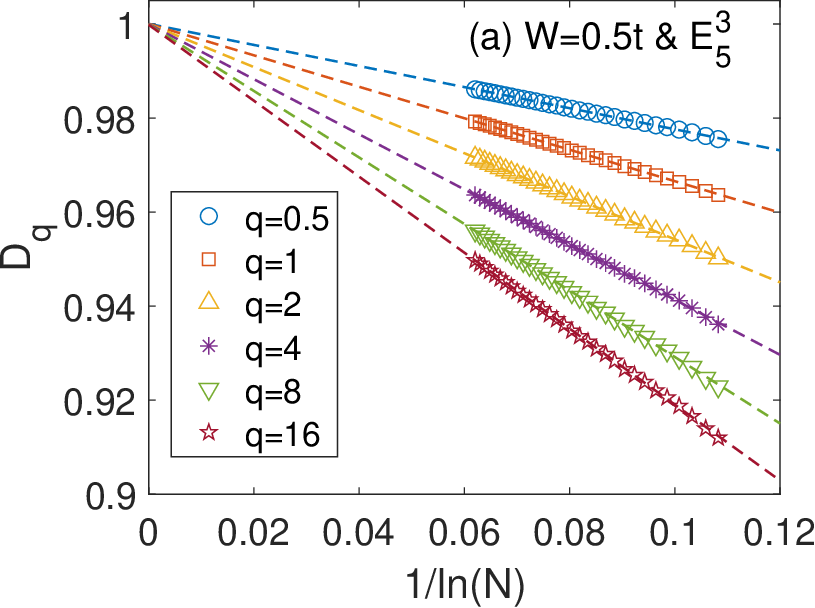}
 \includegraphics[width=1.6in]{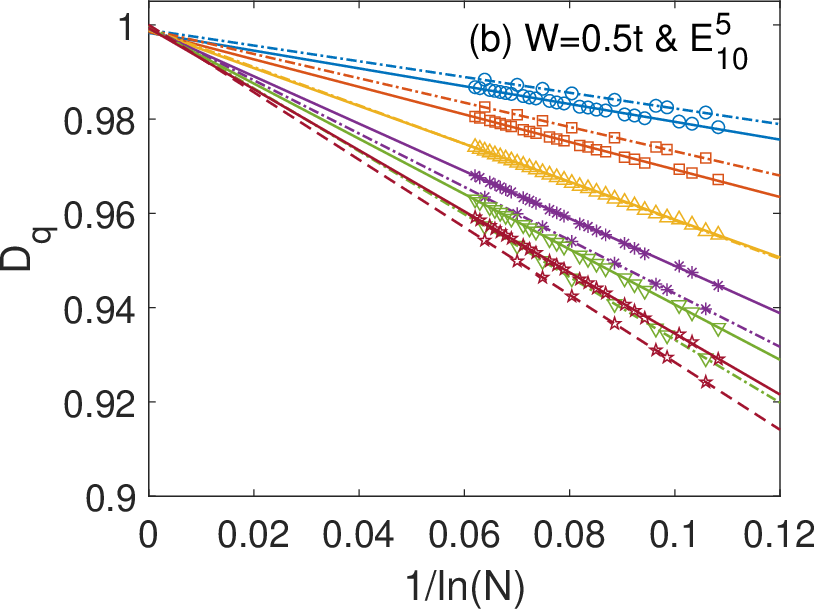}
 \vspace{0.3cm}

 \includegraphics[width=1.6in]{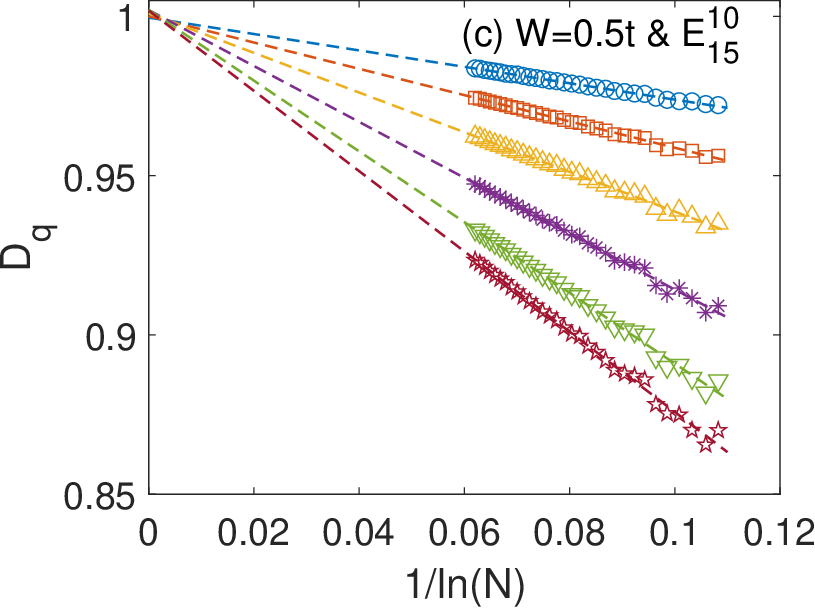}
 \includegraphics[width=1.6in]{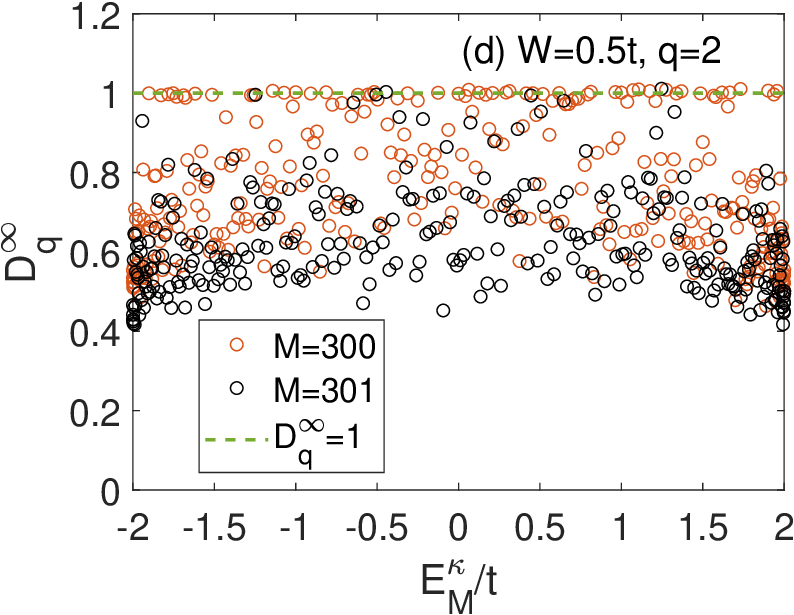}
 \caption{At $W=0.5t$, FDs $D_q$ versus $1/\ln(N)$ for energies $E$ that are nearest to (a) $E^3_5$, (b) $E^5_{10}$ and (c) $E^{10}_{15}$, respectively. (d) At $q=2$, the $D_q^{\infty}$ versus $E^\kappa_M$ with $M=300$ and $301$. The dashed lines in (a)-(c) are linearly fitted to corresponding data. In (a), there are two branches in the curves of $D_q$ versus $1/\ln(N)$. }\label{Fig11}
\end{figure}

%%%%%%%%%%%%%%%%%%%%%%%%
\begin{figure}[!htbp]%fig12
\centering
 \includegraphics[width=1.6in]{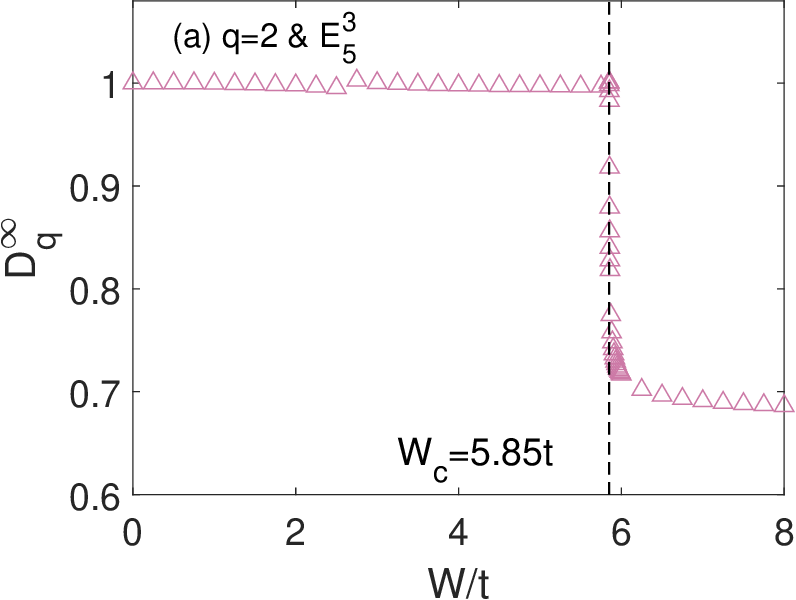}
 \includegraphics[width=1.6in]{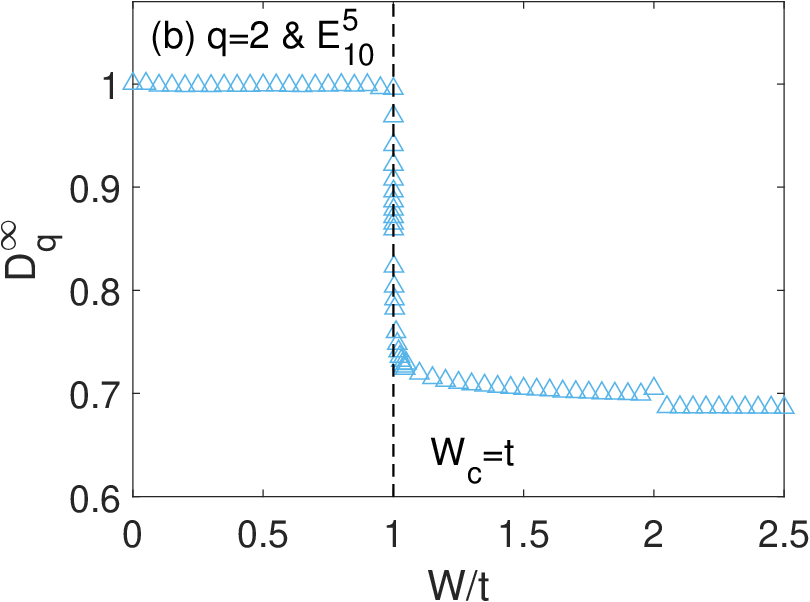}
 \vspace{0.3cm}

 \includegraphics[width=1.6in]{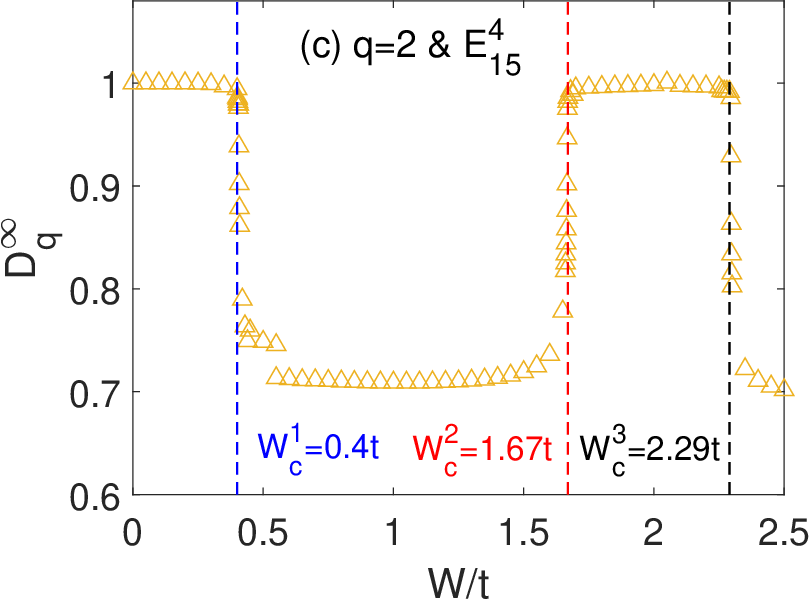}
 \includegraphics[width=1.6in]{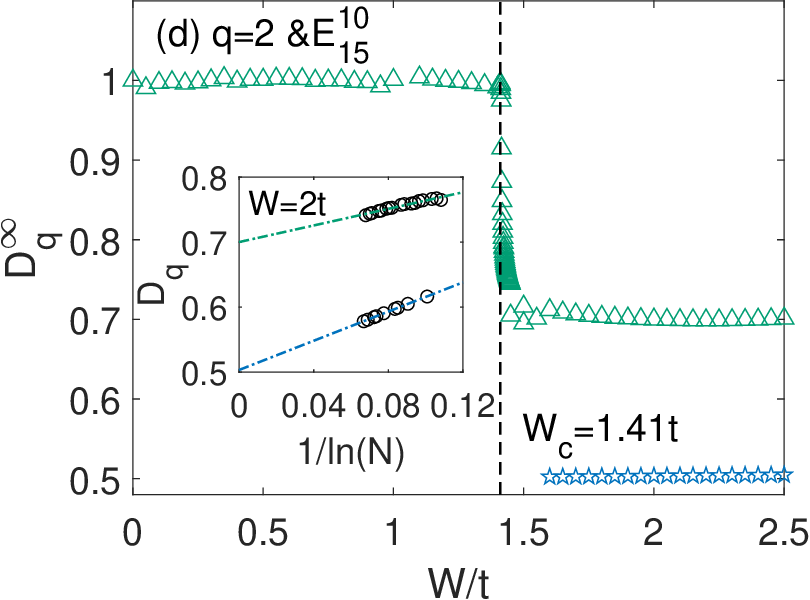}

\caption{At $q=2$, FDs $D_q^{\infty}$ as functions of potential strength $W$ for energies $E$ that are nearest to (a) $E^3_5$, (b) $E^5_{10}$, (c) $E^{4}_{15}$ and (d) $E^{10}_{15}$, respectively.  The vertical dashed lines in (a)-(d) mark the position of $W_c$. The inset in (c) shows $D_q$ versus $1/\ln(N)$ at $W=2t$, where two branches in the curve of $D_q$ versus $1/\ln(N)$.}\label{Fig12}
\end{figure}

At $W=0.5t$, Figs.~\ref{Fig11}(a)-(c) show $D_q$ linearly decrease with $1/\ln(N)$. When $N\to\infty$, $D_q^{\infty}\approx1$, which indicates these states are extended. For $M=300$ and $301$,  at $q=2$, the $D^{\infty}_q$ versus $E^\kappa_M$ are plotted in Fig.~\ref{Fig11}(d). It shows for $M=300$, most of $D_q^{\infty}$ almost equal to ones, which indicates these states are extended. At the same time, many are smaller than ones. In contrast, for $M=301$, almost all $D_q^{\infty}$ are smaller than ones, i.e., they are localized ones.

Fig.~\ref{Fig12} displays the variations of $D_q^{\infty}$ with $W$.
It shows for the four $E^\kappa_M$, there exist regions that $D_q^{\infty}\approx1$ and $D_q^{\infty}<1$. The $W_c$ separates the two types of regions, which agrees with that shown in Fig.~\ref{Fig6} and Fig.~\ref{Fig8}.

\subsection{Effect of randomness}\label{Sec35}
Three kinds of randomness are considered, i.e., disordered on-site potentials, randomly arranged patches and fluctuations in patch sizes.

\subsubsection{Disordered on-site potentials}

\begin{figure}[!htbp]%fig13
\centering
 \includegraphics[width=1.6in]{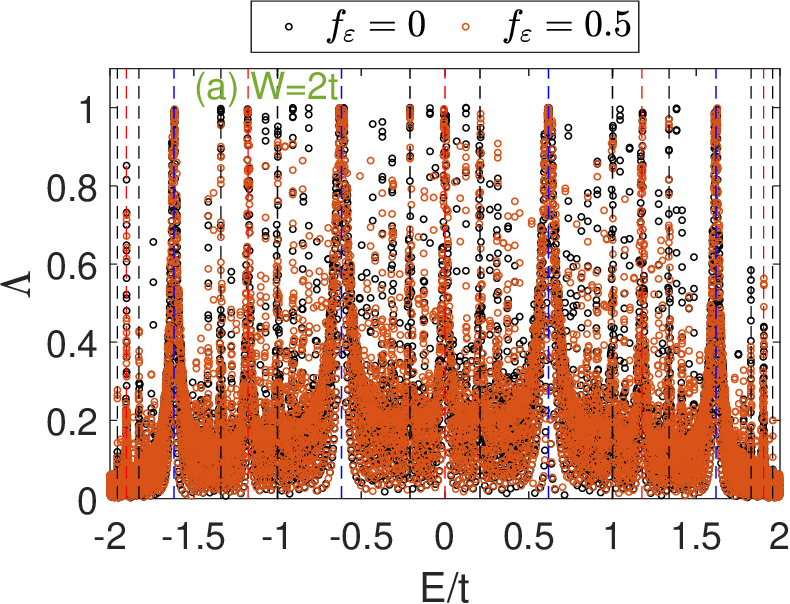}
 \vspace{0.3cm}

 \includegraphics[width=1.6in]{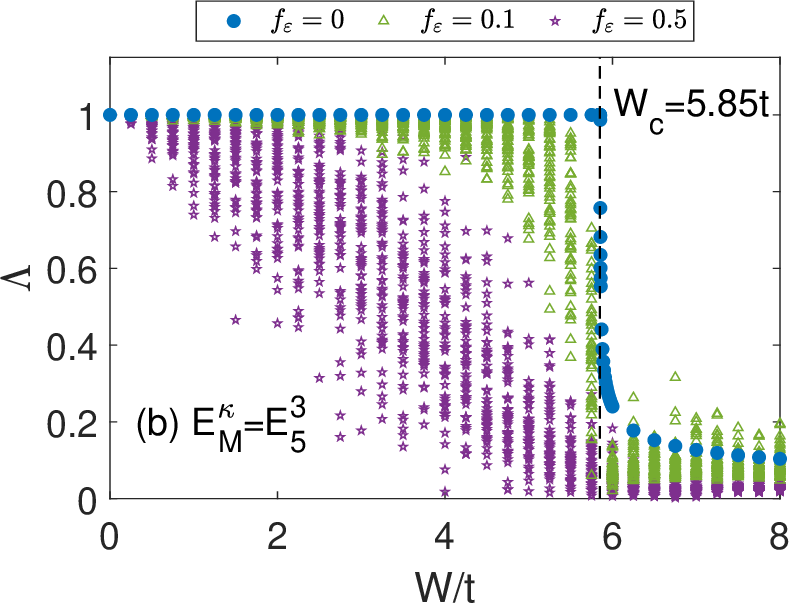}
 \includegraphics[width=1.6in]{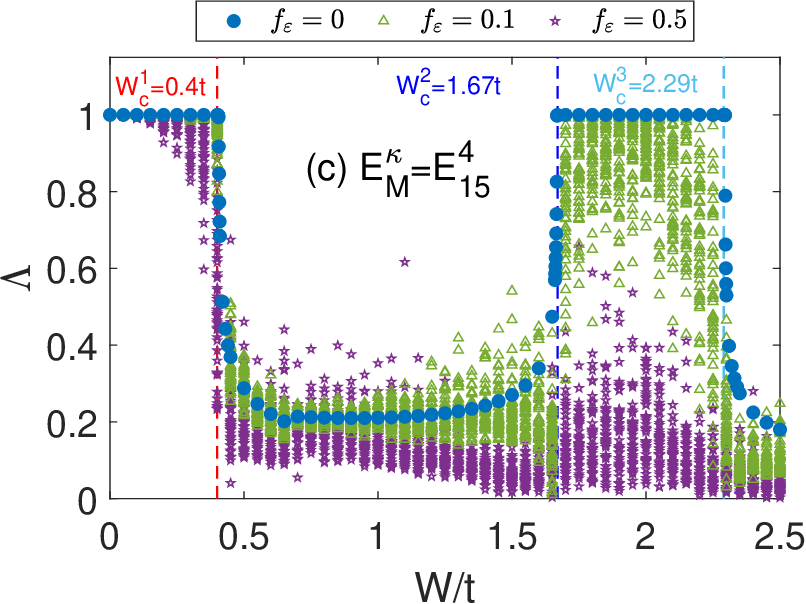}
\caption{(a) RLTs $\Lambda$ as functions of energies $E$ at $W=2t$. The $\Lambda$ as functions of potential strengths $W$ at energies $E$ that are nearest to (b) $E^3_5$ and (c) $E^{4}_{15}$, respectively, where results of $50$ random realizations of $\{\varepsilon_n\}$ are given for each $W$. The vertical dashed lines in (a) are the same in Fig.~\ref{Fig3}. The vertical dashed lines in (b) and (c) mark the position of $W_c$ for $f_{\varepsilon}=0$. System size $N=10,273$ ($j_m=64$) for (a), and $N=[10^6]=1,002,040$ ($j_m=633$) for (b) and (c).}\label{Fig13}
\end{figure}

For the kind of randomness, the on-site potential for B-type sites in Eq.(\ref{EQ2}) becomes $\varepsilon_n=W(1+f_{\varepsilon}\xi_n )$, where $\xi_n$ is a random variable uniformly chosen within the range $[-1/2,1/2]$ and $f_{\varepsilon}$ characterizes the degree of randomness. The patch sizes and their arrangement in space are the same as that in Fig.~\ref{Fig1}.

In Fig.~\ref{Fig13}(a), we plot RLTs $\Lambda$ versus energies $E$ for $f_{\varepsilon}=0.5$, where $W=2t$ is as an example. By contrast, we also plot $\Lambda$ for $f_{\varepsilon}=0$. We find $\Lambda$ will decrease when randomness presents, i.e., disorder can induce localization. However, $\Lambda$ are relatively large when $E$ are around $E^\kappa_M$, which means these state are more extended than other states. Figs.~\ref{Fig13}(b) and (c) show the larger the randomness is, the smaller the  $\Lambda$ is; in general, when randomness presents, $\Lambda$ are relative large in extended-state-W regions [randomness vanishes, seen Fig.~\ref{Fig6}(e), the same below].

\subsubsection{Randomly arranged patches}

\begin{figure}[!htbp]%fig14
\centering
 \includegraphics[width=1.6in]{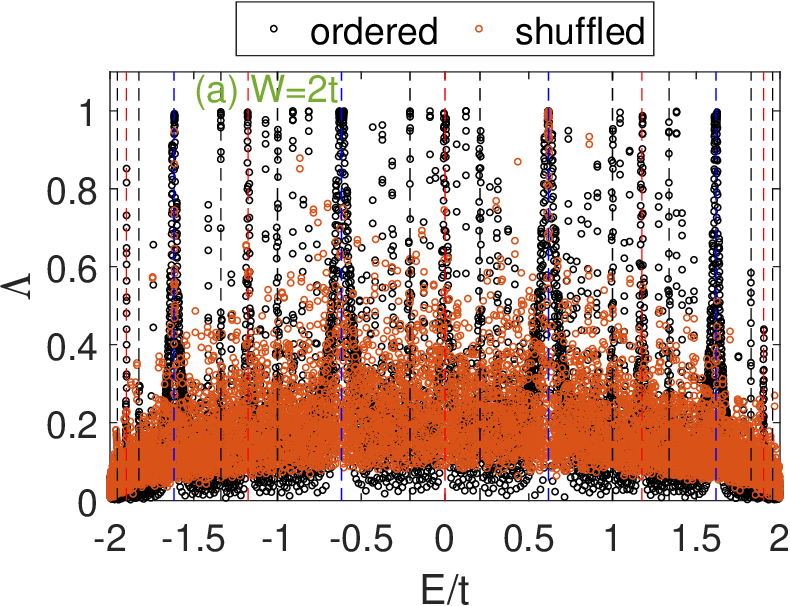}
 \vspace{0.3cm}

 \includegraphics[width=1.6in]{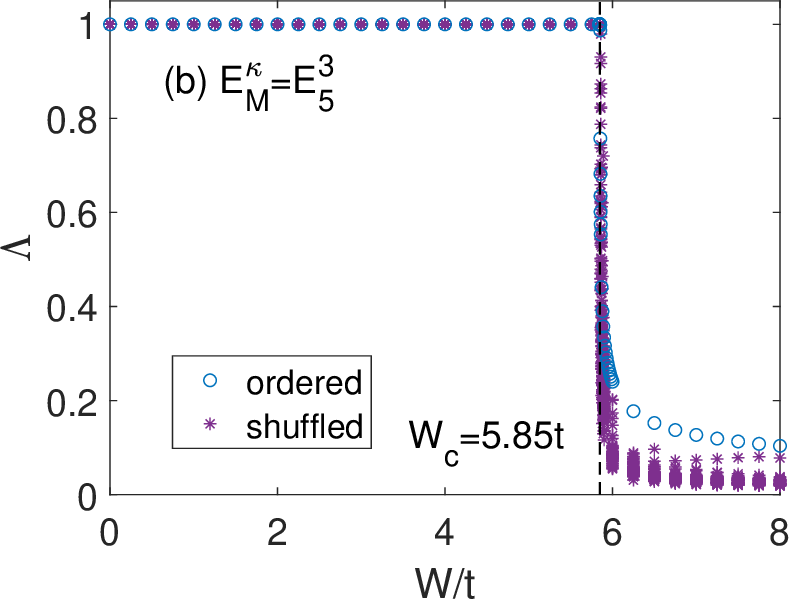}
 \includegraphics[width=1.6in]{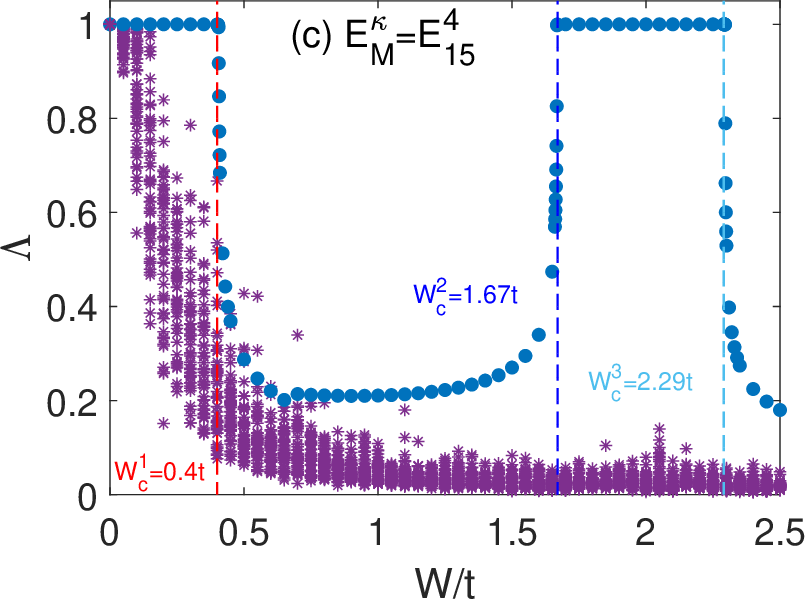}
\caption{(a) RLTs $\Lambda$ as functions of energies $E$ at $W=2t$. The $\Lambda$ as functions of potential strengths $W$ at energies $E$ that are nearest to (b) $E^3_5$ and (c) $E^{4}_{15}$, respectively, where results of $50$ random realizations of $\{\varepsilon_n\}$ are given for each $W$. The vertical dashed lines in (a) are the same in Fig.~\ref{Fig3}. The vertical dashed lines in (b) and (c) mark the position of $W_c$ for the ordered case. System size $N=10,273$ ($j_m=64$) for (a), and $N=[10^6]=1,002,040$ ($j_m=633$) for (b) and (c).}\label{Fig14}
\end{figure}

As mentioned in Sec.\ref{Sec2}, the $j$th patch in Fig.~\ref{Fig1} has $s_j=d_0+(j-1)d$ A-type sites, i.e., $s_j$-mer,  where $j=0,1,\cdots,j_m$. These mers are arranged in order of increasing size. We refer to it as the ordered lattice. In the random-dimer model~\cite{DU90} as well as its variants~\cite{GI93,FA97,EV93,GO16,IZ95}, dimer, trimer, dimer-trimer and $n$-mer present randomly in space. Similarly, we can randomly shuffle all the patches, i.e., \{$s_j$-mer\}, in space. We call it the shuffled lattice. For this kind of randomness,  the on-site potential $\varepsilon_n$ do not change, which are chosen according to Eq.(\ref{EQ2}).

In Fig.~\ref{Fig14}(a), we plot the $\Lambda$ versus energies $E$ for the shuffled lattices at $W=2t$. Compared to that for the ordered ones, $\Lambda$ almost does not change when $E$ are around $E^3_5$ and it rapidly decreases at other $E^\kappa_M$. For $E^3_5$, Fig.~\ref{Fig14}(b) shows $\Lambda$ almost equal to ones in the extended-state-W region, which means these states are extended. In fact, we can get similar results for other $E^\kappa_5$ ($\kappa=1, 2$ and $4$). For other $E^\kappa_M (M=10,15,\cdots)$, the same as shown in Fig.~\ref{Fig14}(c), $\Lambda$ heavily decrease even in extended-state-W regions when randomness presents, i.e., extended states will disappear except at $W=0$.  We know for the shuffled lattices, the size difference between nearest-neighbour patches will be $M'=5m'$ and $m'$ may be $1,2,\cdots$. As mentioned in Sec.~\ref{Sec32}, the corresponding locally-extended localized states have the energies $E^{{\kappa}'}_{M'}=-2\cos\frac{{{\kappa}'}\pi}{5m'}$, i.e., Eq.(\ref{EQ5}). For each patch, there always exists $\kappa'$  satisfying $E^{{\kappa}'}_{M'}=E^\kappa_5$ (resonance conditions). So these states with energies $E^\kappa_5$ can be merged together to form extended states. However, if $M=10,15,\cdots$, the relation that $E^{{\kappa}'}_{M'}=E^\kappa_M$ is not always satisfied, so states with these $E^\kappa_M (M>5)$ are localized ones.

\subsubsection{Fluctuations in patch sizes}

\begin{figure}[!htbp]%fig15
\centering
 \includegraphics[width=1.6in]{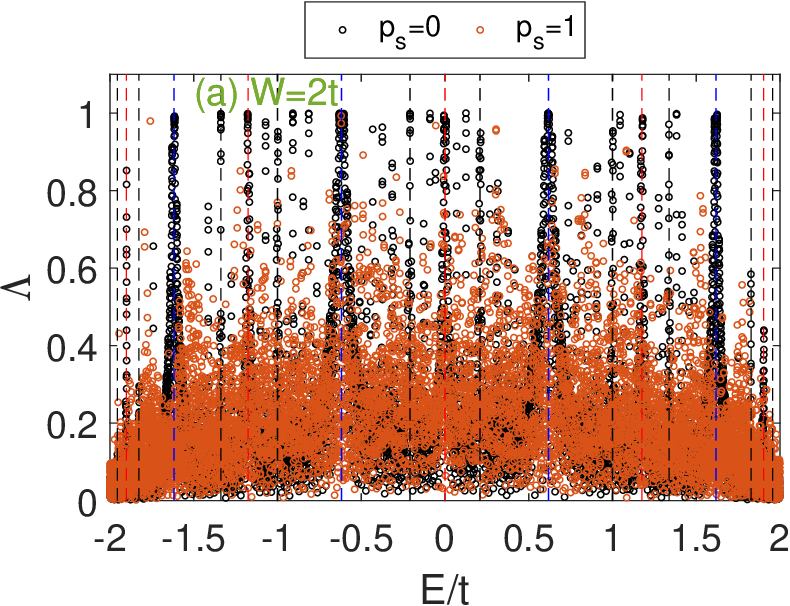}
 \vspace{0.3cm}

 \includegraphics[width=1.6in]{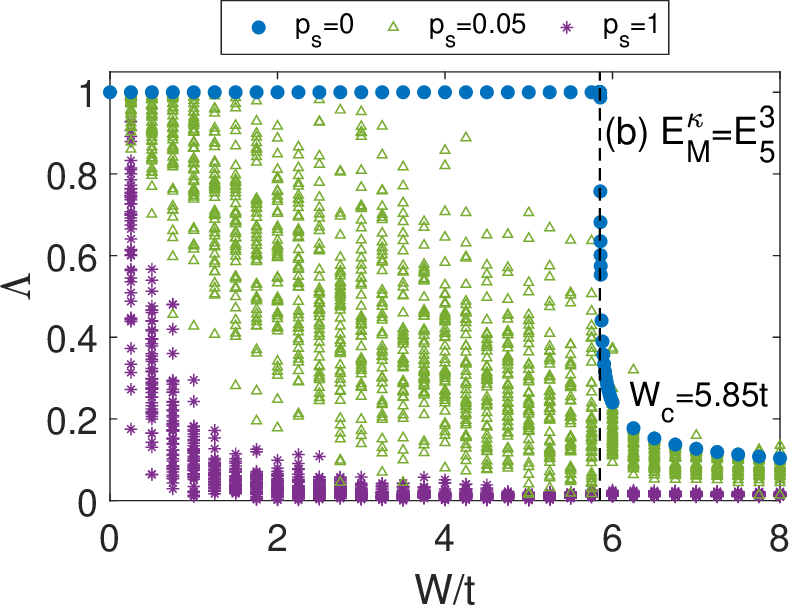}
 \includegraphics[width=1.6in]{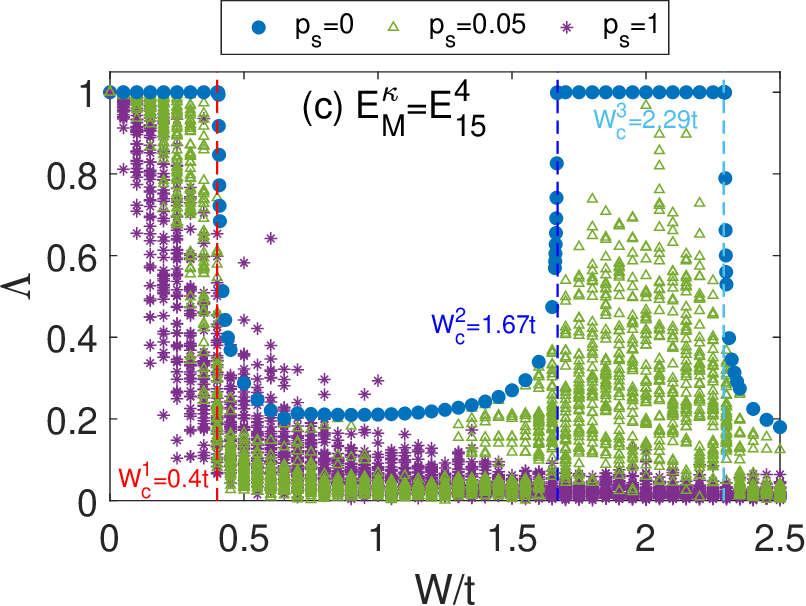}
\caption{(a) RLTs $\Lambda$ as functions of energies $E$ at $W=2t$. The $\Lambda$ as functions of potential strengths $W$ at energies $E$ that are nearest to (b) $E^3_5$ and (c) $E^{4}_{15}$, respectively, where results of $50$ random realizations of $\{\varepsilon_n\}$ are given for each $W$. The vertical dashed lines in (a) are the same in Fig.~\ref{Fig3}. The vertical dashed lines in (b) and (c) mark the position of $W_c$ for $p_{s}=0$. At $p_s=0$, system size $N=10,273$ ($j_m=64$) for (a), and $N=[10^6]=1,002,040$ ($j_m=633$) for (b) and (c); at $p_s=0.05$ and $1$, system sizes approach to just mentioned $N$s, respectively.}\label{Fig15}
\end{figure}

Another randomness is that there are small fluctuations in patch sizes $s_j$ in Fig.~\ref{Fig1}. For the randomness, we consider $s_j=d_0+(j-1)d+[\Delta_{s}\eta_j$] with probability $p_s$, and $s_j=d_0+(j-1)d$ with probability $1-p_s$, where $[Z]$ represents the integer of $Z$, $\eta_j$ is a random variable uniformly chosen within the range $[-1/2,1/2]$. For this kind of randomness, we do not alter the on-site potential $\varepsilon_n$ in Eq.(\ref{EQ2}).

In Fig.~\ref{Fig15}(a), when $W=2t$, we plot the $\Lambda$ versus energies $E$. We take $\Delta_{s}=5$ as an example. Comparing with that for $p_s=0$, we find $\Lambda$ will decrease for $p_s=1$. The $\Lambda$ as functions of potential strengths $W$ are plotted in Figs.~\ref{Fig15}(b) and (c) at energies $E$ that are nearest to $E^3_5$ and  $E^{4}_{15}$, respectively. They show $\Lambda$ become smaller as randomness presents, which means there are absences of extended states. Resonance conditions can not be satisfied for patches with size fluctuations, so all states are localized except at $W=0$.

\section{Conclusions}\label{Sec4}
 A family of 1D aperiodic lattices with linearly varying patches is introduced. Analytically, structure factors show these lattices have strong spatial correlations. In the frame of nearest-neighbour tight-binding models, we show extended states at resonance levels. Three quantities, \textit{i.e.}, local tensions, Lyapunov exponents and fractional dimensions, all can certify the nature of these extended states. These studies may be useful to design high-quality one-frequency selection devices in optoelectronics, optical communication applications and other fields.

%\vspace{0.3cm}
\begin{acknowledgments}
The author would like to thank Hongli Zeng and Yaoxian Zheng for fruitful discussions and useful comments.
This work was supported by the National Natural Science Foundation of China (Grant No. 62375140).\\%Zhaosm
\end{acknowledgments}
\vspace{0.3cm}

\textbf{Appendix A: Scaling laws of structure factors}\\
\begin{table}[tbp]\centering
	\caption{\label{tab}%
		Values of $mod(\widetilde{Z}^{L}_j,2)$.
	}
\label{tab1}
	%\begin{ruledtabular}
		\begin{tabular}{ccccc}
            \hline
            \hline
			%~& $mod(\widetilde{Z}^{L}_j,2)$ & \\
             \multirow{2}{*}{$\ell$}&\multicolumn{4}{c}{$mod(\widetilde{Z}^{L}_j,2)$}\\
             &$L=7$ & $L=9$ & $L=11$ & $L=13$\\
            %$\ell$ &$L=7$ & $L=9$ & $L=11$ & $L=13$\\
			%\colrule
            \hline			
            $0$  & $0$   & $0$   & $0$     & $0$     \\
            $1$  & $6/7$ & $2/3$ & $~6/11$ & $~6/13$ \\
            $2$  & $8/7$ & $4/9$ & $0$     & $22/13$ \\
            $3$  & $6/7$ & $4/3$ & $~4/11$ & $22/13$ \\
            $4$  & $0$   & $4/3$ & $18/11$ & $~6/13$ \\
            $5$  & $4/7$ & $4/9$ & $20/11$ & $0$     \\
            $6$  & $4/7$ & $2/3$ & $10/11$ & $~4/13$ \\
            $7$  &       & $0$   & $10/11$ & $18/13$ \\
            $8$  &       & $4/9$ & $20/11$ & $16/13$ \\
            $9$  &       &       & $18/11$ & $24/13$ \\
            $10$ &       &       & $~4/11$ & $16/13$ \\
            $11$ &       &       &         & $18/13$ \\
            $12$ &       &       &         & $~4/13$ \\
            \hline
            \hline
		\end{tabular}	
    %\end{ruledtabular}
\end{table}

In Fig.1, the position of inlaid B-type site $Z_j=jd_0+j(j-1)d/2+j$ with $j=0,1,2,\cdots,j_m$.
When $d_0=2$ and $d=5$,
\begin{equation}
Z_j=\frac{5j^2+j}{2}. \tag{{A1}} \label{EQA1}
\end{equation}
We set $k=\frac{2}{L}\widetilde{k}\pi$ and $\widetilde{Z}^L_j=\frac{5j^2+j}{L}$, where $L$ is an integer and $\widetilde{k}=1,2,\cdots,[L/2]$. Then
\begin{equation}
kZ_j=\widetilde{k}\pi\widetilde{Z}^L_j. \tag{{A2}} \label{EQA2}
\end{equation}
We represent $j=Lj_0+\ell$ with $\ell=0,1,\cdots,L-1$. At $L=7, 9, 11$ and $13$, the values of $mod(\widetilde{Z}^{L}_j,2)$ are listed in TABLE \ref{tab}.

Based on $mod(\widetilde{Z}^{L}_j,2)$ and the definition of structure factor in Eq.(3), at $j_m\to\infty$, we get
\begin{equation}
S(k)=S_Lj_m^2,\tag{{A3}} \label{EQA3}
\end{equation}
i.e., the scaling parameter $\alpha=2$,
where
%---widetext
\begin{widetext}

\begin{equation}
\begin{aligned}
S_7=\bigg|\frac{1}{7}\{2+2\exp(i\frac{4}{7}\widetilde{k}\pi)+2\exp(i\frac{6}{7}\widetilde{k}\pi)+\exp(i\frac{8}{7}\widetilde{k}\pi)\}\bigg|^2, \label{EQA4}
\end{aligned}\tag{{A4}}
\end{equation}

\begin{equation}
\begin{aligned}
S_9=\bigg|\frac{1}{9}\{2+2\exp(i\frac{2}{3}\widetilde{k}\pi)+2\exp(i\frac{4}{3}\widetilde{k}\pi)+3\exp(i\frac{4}{9}\widetilde{k}\pi)\}\bigg|^2, \label{EQA5}
\end{aligned}\tag{{A5}}
\end{equation}

\begin{equation}
\begin{aligned}
S_{11}=\bigg|\frac{1}{11}\{2+2\exp(i\frac{4}{11}\widetilde{k}\pi)+\exp(i\frac{6}{11}\widetilde{k}\pi)+2\exp(i\frac{10}{11} \widetilde{k}\pi)
+2\exp(i\frac{18}{11} \widetilde{k}\pi)+2\exp(i\frac{20}{11} \widetilde{k}\pi)\}\bigg|^2,\label{EQA6}
\end{aligned}\tag{{A6}}
\end{equation}
and
\begin{equation}
\begin{aligned}
S_{13}=\bigg|\frac{1}{13}\{2+2\exp(i\frac{4}{13}\widetilde{k}\pi)+2\exp(i\frac{6}{13}\widetilde{k}\pi)+2\exp(i\frac{16}{13} \widetilde{k}\pi)
+2\exp(i\frac{18}{13} \widetilde{k}\pi)+2\exp(i\frac{22}{13} \widetilde{k}\pi)+\exp(i\frac{24}{13} \widetilde{k}\pi)\}\bigg|^2.\label{EQA7}
\end{aligned}\tag{{A7}}
\end{equation}

\end{widetext}
%---widetext
For all $\widetilde{k}$, $S_7=1/7$, $S_{11}=1/11$ and $S_{13}=1/13$. At $\widetilde{k}=3$, $S_9=1/3$, and at $\widetilde{k}=1,2$ and $4$, $S_9=1/9$, respectively.\\

%%%%%%%%%%%%%%%%%%%%%%%%%%%
\textbf{Appendix B: Resonance levels}\\

\begin{figure}[!htbp]%fig16
\centering
 \includegraphics[width=2.5in]{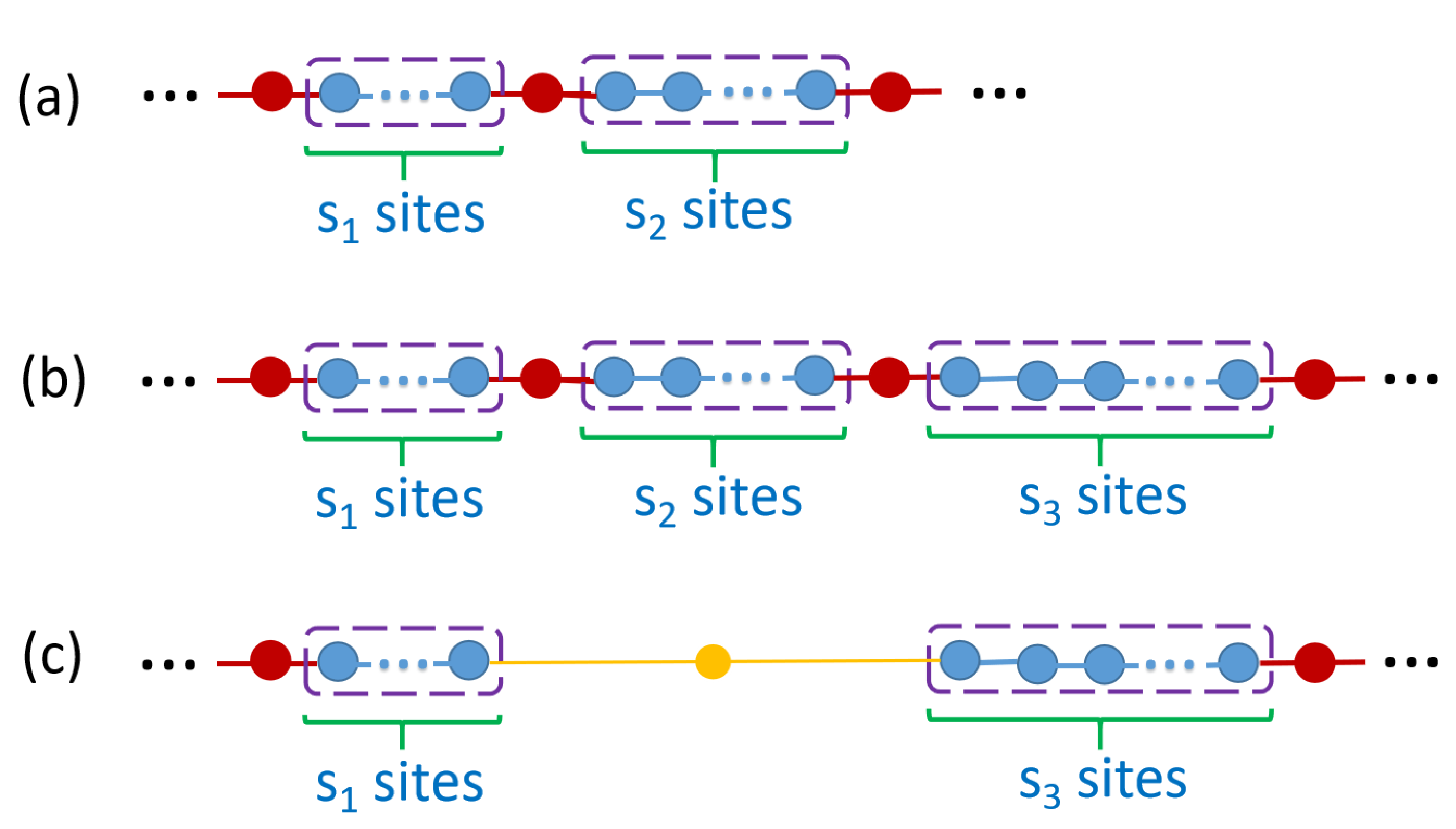}
 \caption{The blue patches have $s_j$ sites and inlaid single red sites link these patches.
 }\label{Fig16}
 \end{figure}

The Schr\"{o}dinger equation for the Hamiltonian in Eq.(1) can be written as
\begin{equation}
-\psi_{n-1}-\psi_{n+1}+\epsilon_n\psi_{n}=E\psi_{n}. \tag{{B1}} \label{EQB1}
\end{equation}
It can be rewritten in terms of the transfer matrix $T(n)$,
\begin{equation}
\Psi_{n+1}=T(n) \Psi_{n}=\left(
  \begin{array}{cc}
    \epsilon_n-E & -1\\
     1           &  0\\
  \end{array}
\right)
\Psi_{n}, \tag{{B2}} \label{EQB2}
\end{equation}
where
\begin{equation}
\Psi_n=\left(
  \begin{array}{c}
    \psi_n\\
    \psi_{n-1}\\
  \end{array}
\right). \tag{{B3}} \label{EQB3}
\end{equation}

We set the matrix $A=\left( \begin{smallmatrix} -E & -1\\ 1 & 0 \end{smallmatrix} \right) $  and
$B=\left( \begin{smallmatrix}     W-E & -1\\ 1 & 0 \end{smallmatrix} \right) $, which corresponds to blue (A-type) sites and red (B-type) sites in Fig.{\ref{Fig16}}. We consider a unit, which includes two patches and two inlaid sites [seen Fig.{\ref{Fig16}} (a), not including the most right red (B-type) site], so there are $s_1+s_2+2$ sites. The total transfer matrix
\begin{equation}
T=BA^{s_2}BA^{s_1}. \tag{{B4}} \label{EQB4}
\end{equation}
Using the spectral decomposition method, $A=U\Lambda{U^{-1}}$, where $\Lambda=\left( \begin{smallmatrix} \lambda_{+} & 0\\ 0 & \lambda_{-} \end{smallmatrix} \right)$, and $U$'s first and second rows are the eigenvectors of $A$ with eigenvalues $\lambda_{+}=\frac{-E+\sqrt{E^2-4}}{2}$ and  $\lambda_{-}=\frac{-E-\sqrt{E^2-4}}{2}$, respectively. When $|E|<2$, $\lambda_{\pm}=e^{\pm i\theta}$ with that $\sin\theta=\frac{\sqrt{4-E^2}}{2}$
and $\cos\theta=\frac{-E}{2}$. So
\begin{equation}
E=-2\cos\theta \tag{{B5}} \label{EQB5}
\end{equation}
and
\begin{equation}
U=\left(
  \begin{array}{cc}
    e^{i\theta} & 1\\
    e^{-i\theta}&  1\\
  \end{array}
\right), \tag{{B6}} \label{EQB6}
\end{equation}
where $i=\sqrt{-1}$. The trace of total transfer matrix $T$ is

%%%%%%%%%%%%%%%%%%%%%%%%>>widetex
\begin{widetext}
\begin{equation}
\chi=Tr{(T)}=Tr({B}U\Lambda^{s_2}{U^{-1}}{B}U\Lambda^{s_1}{U^{-1}})=a(W+2cos\theta)^2+b(W+2cos\theta)+c,\tag{{B7}} \label{EQB7}
\end{equation}
where
\begin{equation}
a=\frac{1}{2sin^2\theta}\big\{\cos[(s_1-s_2)\theta]-\cos[(s_1+s_2+2)\theta]\big\},\tag{{B8}} \label{EQB8}
\end{equation}
\begin{equation}
b=-\frac{2}{sin^2\theta}\big\{\cos{\theta}\cos[(s_1-s_2)\theta]-\cos[(s_1+s_2+1)\theta]\big\},\tag{{B9}} \label{EQB9}
\end{equation}
and
\begin{equation}
c=\frac{2}{sin^2\theta}\big\{\cos^2{\theta}\cos[(s_1-s_2)\theta]-\cos[(s_1+s_2)\theta]\big\}.\tag{{B10}} \label{EQB10}
\end{equation}
\end{widetext}
%%%%%%%%%%%%%%%%%%%%%%%%<<widetex
Based on the theory of trace map of transfer matrices~\cite{KO83}, for allowed energies
\begin{equation}
|\chi(E)|\leq2. \tag{{B11}}\label{EQB11}
\end{equation}
States are extended and critical when $|\chi|<2$ and $|\chi|=2$, respectively.

For Eq.(\ref{EQB7}), we consider the condition that
\begin{equation}
\cos[(s_1-s_2)\theta]\approx\cos[(s_1+s_2+2)\theta], \tag{{B12}}\label{EQB12}
\end{equation}
i.e., $a\approx0$ in Eq.(\ref{EQB8}). We replace $\cos[(s_1-s_2)\theta]$ by $\cos[(s_1+s_2+2)\theta]$ in Eqs.(\ref{EQB9}) and (\ref{EQB10}), then
\begin{equation}
b{\approx}\frac{2}{\sin\theta}\sin[(s_1+s_2+2)\theta], \tag{{B13}}\label{EQB13}
\end{equation}
and
\begin{equation}
c{\approx}2\cos[(s_1+s_2+2)\theta]-4\cot{\theta}\sin[(s_1+s_2+2)\theta]. \tag{{B14}}\label{EQB14}
\end{equation}
If $E$ in Eq.(\ref{EQB5}) is represented by
\begin{equation}
E=-2\cos(\theta)=-2\cos(\frac{\widetilde{\kappa}\pi}{s_1+s_2+2}),\tag{{B15}}\label{EQB15}
\end{equation}
$\cos[(s_1+s_2+2)\theta]=\pm1$ and $\sin[(s_1+s_2+2)\theta]=0$, where $\widetilde{\kappa}=1,2,\cdots$. From Eqs.(\ref{EQB13})-(\ref{EQB14}), we get $b\approx0$ and $c\approx\pm2$,
so in Eq.(\ref{EQB7})
\begin{equation}
\chi\approx\pm2, \tag{{B16}}\label{EQB16}
\end{equation}
which is nearly independent of potential strength $W$. In combination with Eq.(\ref{EQB15}),
the condition in Eq.(\ref{EQB12}) indicates
\begin{equation}
E\approx E^\kappa_M=-2\cos(\frac{\kappa\pi}{M}) \tag{{B17}}\label{EQB17}
\end{equation}
with $M=|s_1-s_2|$ and $\kappa=1,2,\cdots,M-1$. The corresponding $\chi\approx\pm2$, and states are extended or critical. If $s_1=s_2=s$, according to Eqs.(\ref{EQB12}) and (\ref{EQB15}),
we get
\begin{equation}
E^\kappa_M=-2\cos(\frac{\kappa\pi}{s+1}), \tag{{B18}}\label{EQB18}
\end{equation}
which agrees with the Bloch's theory.

\begin{figure}[!htbp]%fig17
 \includegraphics[width=1.6in]{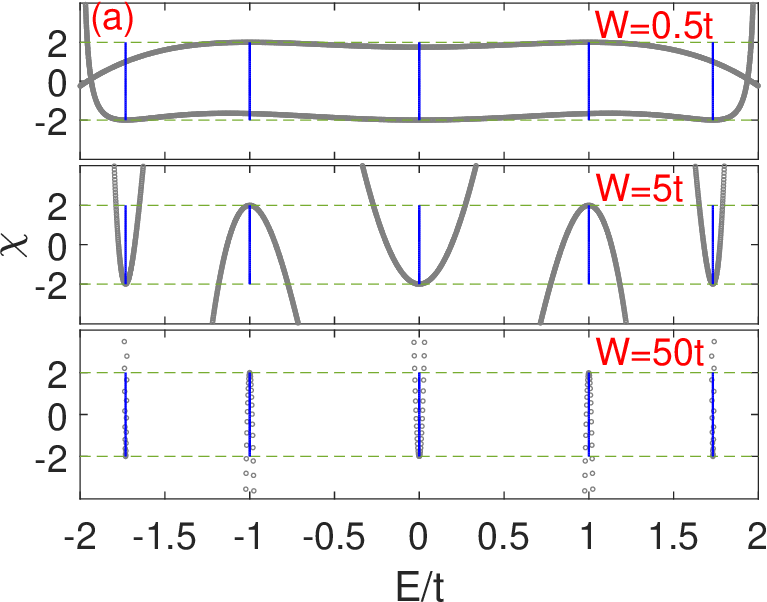}
 ~~~\includegraphics[width=1.65in]{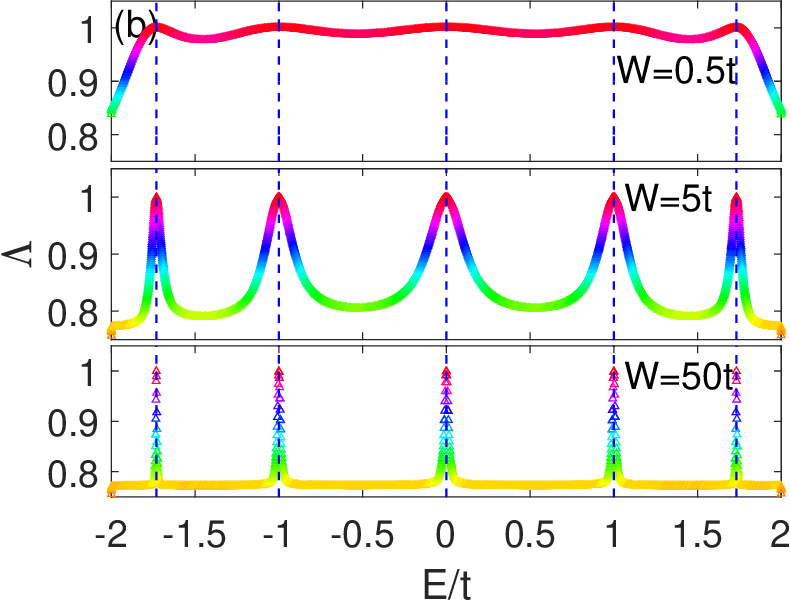}
 \caption{(a) The trace $\chi$ and (b) RLTs $\Lambda$ as functions of energies $E$ at potential strengths $W=0.5t, 5t$ and $50t$, respectively. Here $s_1=3330$ and $s_2=3336$. The blue vertical lines mark the positions $E^\kappa_M$ with $M=|s_1-s_2|=6$.}\label{Fig17}
\end{figure}

Based on Eq.(\ref{EQB7}), we plot $\chi$ versus energies $E$ in Fig.\ref{Fig17}(a) at $W=0.5t, 5t$ and $50t$, respectively. It shows $\chi\approx\pm2$ when $E$ are at $E^\kappa_M$, which agrees with theoretical conclusions. The reduced local tensions (RLTs) $\Lambda$ can directly characterize state localization properties~\cite{DE19}. Fig.\ref{Fig17}(b) shows $\Lambda\to1$ when $E$ are at $E^\kappa_M$, which indicates these states are extended.

%---------------
\begin{figure}[!htbp]%fig18
 \includegraphics[width=1.6in]{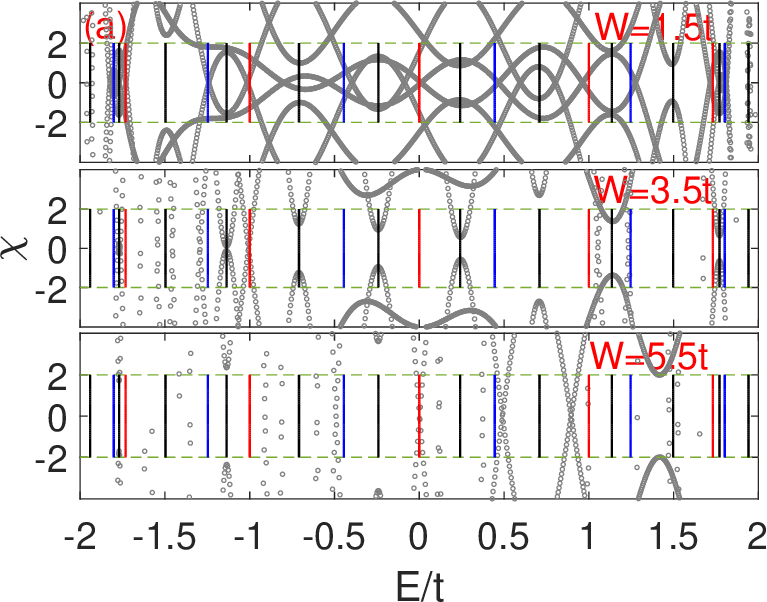}
 ~~~\includegraphics[width=1.65in]{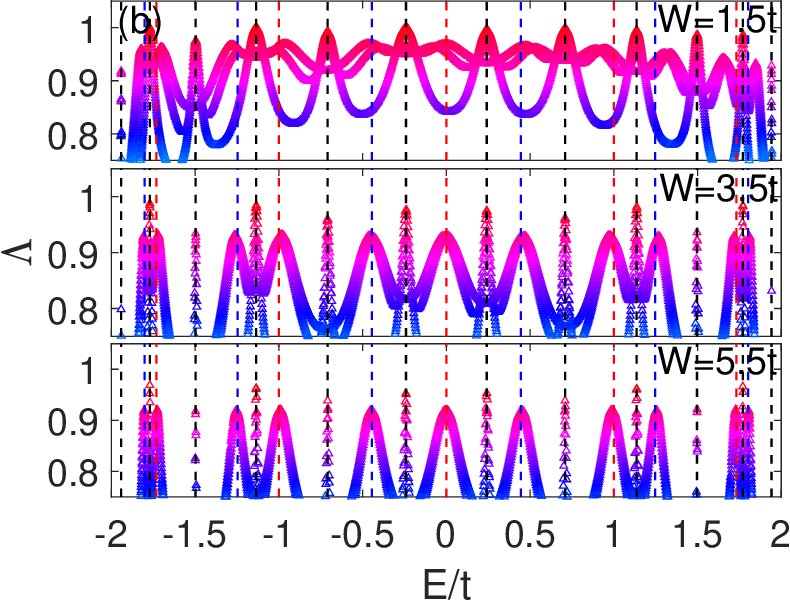}
 \caption{(a) The trace $\chi$ and (b) RLTs $\Lambda$ as functions of energies $E$ at potential strengths $W=1.5t, 3.5t$ and $5.5t$, respectively. Here $s_1=2331$, $s_2=2338$ and $s_3=2344$. The vertical lines mark the positions $E^\kappa_M$ with $M=|s_1-s_2|=7$ (blue), $M=|s_2-s_3|=6$ (red) and $M=|s_1-s_3|=13$ (black), respectively.}\label{Fig18}
\end{figure}

Then, we consider a unit which consists of three patches and three inlaid sites [Fig.{\ref{Fig16}} (b), not including the most right red (B-type) site]. Using the numerically accurate renormalization scheme~\cite{FA92}, both the sites in the intermediate patch and the intermediate inlaid sites can be renormalized into ``one'' inlaid' site [the yellow site in Fig.{\ref{Fig16}} (c)], so they can be taken as ``two patches''. Eq.(\ref{EQB17}) also holds but $M$ is the size difference of the patches at two edges. Based on Eq.(\ref{EQB2}), we directly calculate $\chi$. We plot $\chi$ and $\Lambda$ in Figs.{\ref{Fig18}}(a) and (b), respectively. It shows generally, $\chi$ are relative small and $\Lambda$ are relative large when $E$ are around $E^\kappa_M$. At some $E^\kappa_M$, $\Lambda\to1$, which indicates these states are extended. Similarly, for more patches, the results are the same, but the renormalized ``inlaid'' site may induce localized effects. For a few of patches (we call it a super-patch), there are states with energies $E^\kappa_M$.  When these super-patches are linked together by inlaid B-type sites, the energies of whole lattices around $E^\kappa_M$ may become resonance levels if they are allowed energies, and related states may be extended.\\

\textbf{Appendix C: Local tensions at different $d$}\\

\begin{figure}[!htbp]%fig19
\centering
  \includegraphics[width=1.6in]{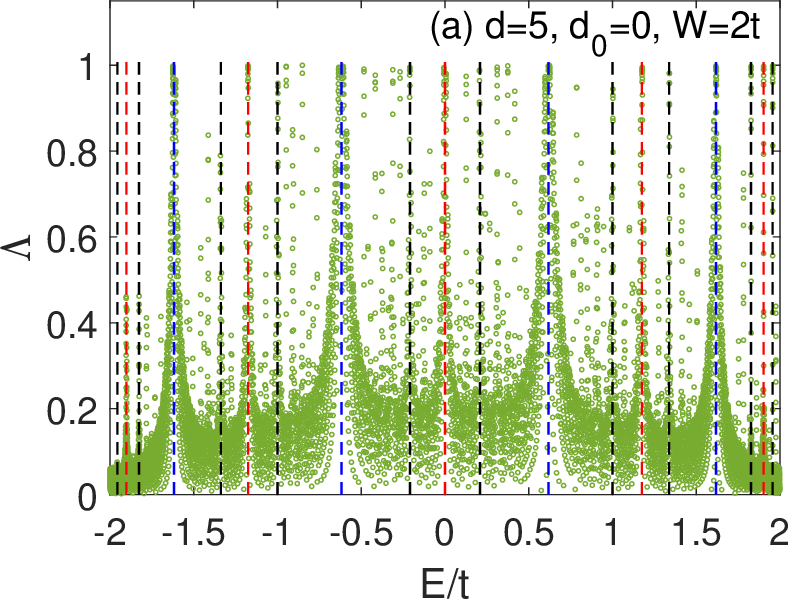}
  \includegraphics[width=1.6in]{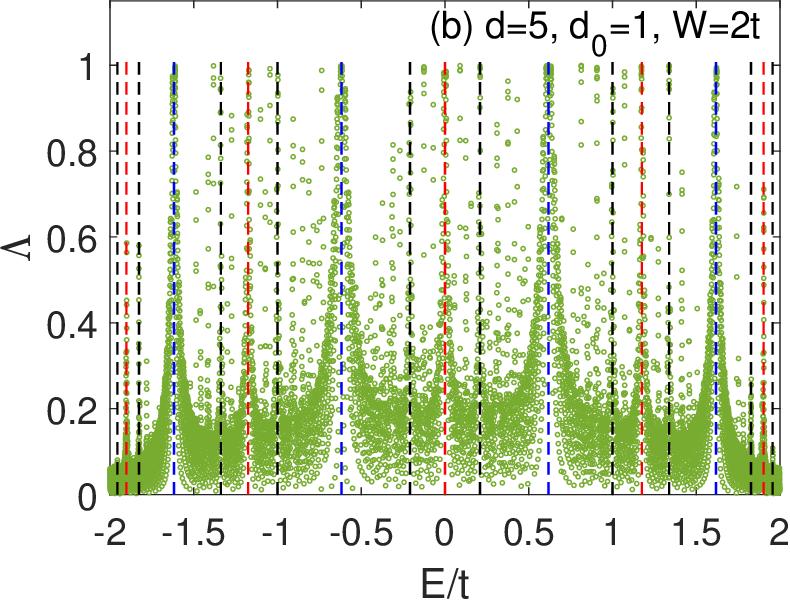}
 \vspace{0.2cm}

  \includegraphics[width=1.6in]{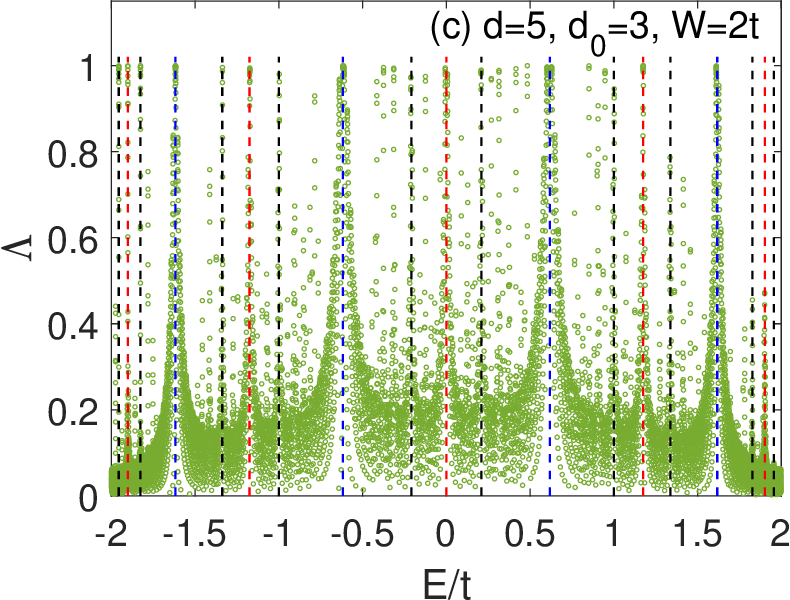}
  \includegraphics[width=1.6in]{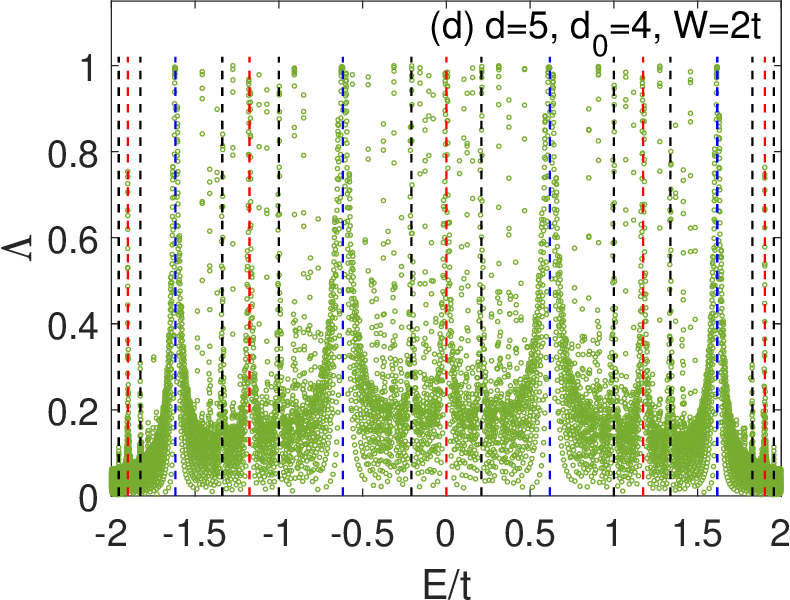}
 \caption{At $d=5$ and $W=2t$, the RLTs $\Lambda$ versus energies $E$ for (a) $d_0=0$, (b) $d_0=1$, (c) $d_0=3$ and (d) $d_0=4$, respectively. The vertical dashed lines mark the positions $E^\kappa_M$ with $M=5$ (blue),  $M=10$ (blue), and  $M=15$ (black). System sizes $N=[10^4]$.}\label{Fig19}
\end{figure}

%\begin{figure}[!htbp]%Fig20
%\centering
%  \includegraphics[width=1.6in]{F20A.eps}
%  \includegraphics[width=1.6in]{F20B.eps}
% \vspace{0.2cm}

%  \includegraphics[width=1.6in]{F20C.eps}
%  \includegraphics[width=1.6in]{F20D.eps}
% \caption{At $d_0=2$, the RLTs $\Lambda$ versus energies $E$ for (a) $d=1, W=t$, (b) $d=2, W=2t$, (c) $d=13, W=3t$ and (d) $d=24, W=3t$, respectively. The %vertical dashed lines mark the positions $E^\kappa_M$. In (a), $M=2d$ (blue), $M=3d$ (red) and $M=4d,5d,\cdots,12d$ (black), where $d=1$; in (b), $M=d$ (blue), %$M=2d$ (red) and $M=3d, 4d,\cdots,10d$ (black), where $d=2$; in (c), $M=d$ (blue), $M=2d$ (red) and $M=3d$ (black),  where $d=13$; in (d), $M=d$ (blue), $M=2d$ %(red) and $M=3d$ (black), where $d=24$. System sizes $N=[10^4]$.}\label{Fig20}
%\end{figure}
\FloatBarrier

We know the reduced local tensions (RLTs) $\Lambda$ can directly characterize state localization properties~\cite{DE19}. The larger the $\Lambda$ are, the states are more extended. In Fig.~\ref{Fig19}, at $W=2t$, we plot $\Lambda$ versus energies $E$ at $d_0=0, 1, 3, 4$ with $d=5$, respectively. %The $\Lambda$ at $d_0=2$ %has shown in Fig.~\ref{Fig4}. In Fig.~\ref{Fig20}, at $d_0=2$, we plot $\Lambda$ versus $E$ at $d=1, 2, 13$ and $24$. 
All the figures show $\Lambda$ are relative large when $E$ are around $E^\kappa_M$, which indicates these states are extended (delocalized).\\

\end{document}